\documentclass[fleqn,usenatbib]{mnras}

\usepackage{newtxtext,newtxmath}

\usepackage[T1]{fontenc}

\DeclareRobustCommand{\VAN}[3]{#2}
\let\VANthebibliography\thebibliography
\def\thebibliography{\DeclareRobustCommand{\VAN}[3]{##3}\VANthebibliography}

\usepackage{graphicx}	
\usepackage{amsmath}	
\usepackage{pgfplotstable}
\pgfplotsset{compat=1.18}
\usepackage{booktabs}
\usepackage{multirow}
\usepackage{threeparttable}
\usepackage{caption}
\usepackage{pdflscape}
\usepackage{makecell}

\newcommand{\ms}{$M_{\odot}$}
\newcommand{\swift}{\textit{Swift}}
\newcommand{\maxi}{\textit{MAXI}}
\newcommand{\xspec}{\textit{XSPEC}}

\newcommand{\nicer}{\textit{NICER}}

\newcommand{\rxte}{\textit{RXTE}}

\title[Estimating Distances to BH-LMXBs]{A Dependable Distance Estimator to Black Hole Low-Mass X-ray Binaries}

\author[Abdulghani et al.]{
Y. Abdulghani,$^{1}$\thanks{E-mail: youssefabdulghani@montana.edu}
A. M. Lohfink,$^{1}$
J. Chauhan$^{2, 1}$
\\
$^{1}$Department of Physics, Montana State University, P.O. Box 173840, Bozeman, MT 59717-3840, USA\\
$^{2}$Inter-University Center for Astronomy and Astrophysics, Ganeshkhind, Pune 411007, India\\
}

\date{Accepted XXX. Received YYY; in original form ZZZ}

\pubyear{2024}

\begin{document}
\label{firstpage}
\pagerange{\pageref{firstpage}--\pageref{lastpage}}
\maketitle

\begin{abstract}
Black Hole Low Mass X-ray Binaries (BH-LMXBs) are excellent observational laboratories for studying many open questions in accretion physics. However, determining the physical properties of BH-LMXBs necessitates knowing their distances. With the increased discovery rate of BH-LMXBs, many canonical methods cannot produce accurate distance estimates at the desired pace. In this study, we develop a versatile statistical framework to obtain robust distance estimates soon after discovery. Our framework builds on previous methods where the soft spectral state and the soft-to-hard spectral state transitions, typically present in an outbursting BH-LMXB, are used to place constraints on mass and distance. We further develop the traditional framework by incorporating general relativistic corrections, accounting for spectral/physical parameter uncertainties, and employing assumptions grounded in current theoretical and observational knowledge. We tested our framework by analyzing a sample of 50 BH-LMXB sources using X-ray spectral data from the \swift/XRT, \maxi/GSC, and \rxte/PCA missions. By modeling their spectra, we applied our framework to 26 sources from the 50. Comparison of our estimated distances to previous distance estimates indicates that our findings are dependable and in agreement with the accurate estimates obtained through parallax and H\,{\sc i} absorption methods. Investigating the accuracy of our constraints, we have found that estimates obtained using both the soft and transition spectral information have a median uncertainty (1$\sigma$) of 20\%, while estimates obtained using only the soft spectral state spectrum have a median uncertainty (1$\sigma$) of around 50\%. Furthermore, we have found no instrument-specific biases.
\end{abstract}

\begin{keywords}
accretion, accretion discs – methods: statistical – software: data analysis – stars: black holes – stars: distances – X-rays: binaries
\end{keywords}


	
\section{Introduction}

It is unequivocal that the accretion mechanisms around black holes are not yet fully understood. Black hole X-ray binaries (BHXBs) provide ideal observational laboratories to study the accretion process. In particular, transient low mass X-ray binaries (BH-LMXBs) show variations in their accretion rate at a time scale of days to a few years \citep{fender2016, tetarenko2016}. This renders investigations into the underlying physics of accretion much more feasible compared to supermassive black hole accretion systems present in Active Galactic Nuclei \citep[AGN;][]{Silk1998}.

Depending on the mass of the companion star and the variability of the emission, BHXBs are divided into (1) Persistent, typically high-mass X-ray binaries (BH-HMXBs) where the companion star mass is comparable to or greater than the primary BH mass and mass is always being transferred, (2) Transient, which are found to be mostly low mass X-ray binaries (BH-LMXBs), where the companion star mass is $\le$ \ms~and the mass accretion rate onto the black hole is typically low. However, these transient systems undergo recurrent outbursts in which an increased accretion rate gives rise to a bright emission followed by an exponential decay back to their usual faint state, typically known as quiescence \citep{corral-santana2016, tetarenko2016}. 

There are, however, notable exceptions to this picture of black hole X-ray binaries, one such exception is 4U 1957+11. This source has been studied multiple times since its discovery in 1974 \citep{giacconi1974}. Evidence from most studies strongly suggests that it is a persistent BH-LMXB observed to be in the soft state \citep[e.g.,][]{wijnands2002,gomez2015,barillier2023}

The transient behavior of BH-LMXBs is thought to be driven by instabilities in the accretion disk \citep{Dubus2001, remillard2006} leading to large fluctuations of the accretion rate. During a typical outburst, a rapid increase in the accretion rate (timescale of days to months) causes characteristic changes to the accretion geometry and emitted radiation \citep[e.g.][]{remillard2006,fender2004}.

Furthermore, transient BH-LMXBs are observed to evolve through a series of different spectral states during typical outbursts. This evolution of spectral states is thought to be explained by a hot inner advection-dominated accretion flow (ADAF) and a truncated accretion disk \citep{esin1997}. In this picture, the accretion rate is low initially before the outburst. During that time, the accretion disk does not extend all the way to the black hole but is recessed. As a result, the radiation emitted by the accretion disk during this time is weak.  At low accretion rates, the ``hard" X-ray emission from the inner ADAF dominates over the disk emission, and this state of the BH-LMXB is therefore referred to as the \textit{hard state}. As the accretion rate increases, the accretion disk starts to move closer to the black hole. The disk contribution to the X-ray spectrum increases. The BH-LMXB moves into the \textit{intermediate} state. Then, as the accretion rate gets closer to the maximum possible rate, the disk extends until the innermost stable orbit (ISCO), and the accretion disk spectrum completely dominates the X-ray spectrum. This high luminosity state is called the \textit{soft state}, where the low energy (soft X-ray) disk black body emission dominates. Finally, as the episodic increased accretion event comes to an end, the accreting BH-LMXB fades via the intermediate state back into the quiescence \citep[for review see][and reference therin]{done2007}. Thus, in this framework, the observed radiation is produced primarily via two processes: 1) the thermal emission from the accretion disk itself \citep{shakura1973,mitsuda1984} and 2) the up-scattered emission from the inner ADAF. This up-scattered emission is created by some of the primary disk photons scattering off the hot electrons in the inner ADAF and gaining energy, resulting in a power-law-like spectrum towards higher (harder) X-ray energies \citep{thorne1975}. 

Although the current models for accretion can explain some of the observed behaviour of BH-LMXBs, much remains unknown. Moreover, intrinsic observables such as accretion geometry, jet power, luminosity, and magnetic field strength are needed to refine existing frameworks, such as the inner ADAF and the truncated disk model, and to test new theoretical models. Crucially, getting constraints on such intrinsic properties is often hampered by the poor quality of the distance estimate of these systems.

The most accurate distances to BH-LMXBs can obtained from a model-independent parallax measurement to the companion star/jet \citep[e.g.][]{miller-jones2009,gandhi2019,atri2020,arnason2021}. However, this approach cannot be used if the companion star/jet is too faint, too far or there are not enough observations available to permit the parallax calculation. Other methods include using interstellar extinction \citep{jonker2004}, analyzing the spectrum of the companion star \citep{jonker2004}, using the 21-cm line of neutral hydrogen \citep[e.g.][]{dhawan2000,chauhan2021a}, modelling the X-ray dust scattering halo radial profiles \citep[e.g.][]{trumper1973,thompson2009,kalemci2018}, or measuring the proper motion of jet ejecta \citep[e.g.][]{mirabel1994}. All these methods also suffer from limitations, such as the requirement that the companion star is bright enough for spectral analysis, the existence of bipolar jets and/or that the inclination of the source is known. Their constraints also often suffer from high uncertainties \citep{jonker2004}.

As of December 2023, the catalogue of stellar-mass black holes in X-ray binaries
\footnote{\url{https://www.astro.puc.cl/BlackCAT/index.php}} produced by \citet{corral-santana2016} (\textit{BlackCAT}), lists a total of 70 observed transient sources where the compact object is either a dynamically confirmed BH or a strong BH candidate. From these 70 sources, only 28 have good distance estimates, and 14 have poorly constrained distances (error $> 50 \%$ or no lower/upper bounds). While 28 have \textit{no distance estimate} to the best of our knowledge. It is estimated that the number of BH-LMXBs in our Galaxy is of the order of thousands \citep[e.g.][]{romani1998,kiel2006}. Moreover, there has been an exponential increase in the detection of transient sources in the past two decades \citep{corral-santana2016}, driven by recent missions/instruments such as the Rossi X-ray Timing Explorer \citep[\rxte;][]{rxte}, Neil Gehrels Swift Observatory/Burst Alert Telescope \citep[\swift/BAT;][]{swift_bat}~and the Monitor of All-sky X-ray Image \citep[\maxi;][]{maxi}. This rate is only expected to keep increasing. Keeping up with the ever-growing BH-LMXBs discoveries will require a simple and generally applicable distance estimation method. 

While various BH-LMXB systems (e.g., GRO J0422+32, XTE J1118+480, some outbursts of GX 339-4) have been found to exhibit only the hard state \citep{alabarta2021}, the majority of the current sample of BH-LMXBs (about 70\% of the observed outbursts, according to \citealt{alabarta2021}) are detected in the X-ray band when they undergo a "canonical" outburst. Here, we define a "canonical" outburst as one where the hard, soft, and intermediate states are observed. Consequently, a strong candidate for a distance estimation method is one that utilizes the modeling results of the X-ray spectra from specific spectral states typically present in any canonical BH-LMXB outburst, namely, the soft state and the soft-to-hard transition \citep[e.g.,][]{remillard2006,kalemci2022}. Moreover, throughout the transient BH-LMXB literature, this method has been previously used \cite[e.g.][]{nakahira2012,nakahira2014,oda2019,tominaga2020} however, no systematic study of the robustness of this method was ever conducted.

In this study, we further develop this method, quantify its uncertainties, and implement it in a framework that allows for its application to any BH-LMXB exhibiting the typical outburst pattern. Our paper is organized as follows: Sections~\ref{sec:dataselect} and~\ref{sec:datared} describe how we selected the sample, the spectral data, and its analysis. Subsequently, in Section~\ref{sec:spec_model}, we detail the spectral models used and the results of modeling the spectral data. In Section~\ref{sec:method-overall}, we establish the statistical framework used to obtain our distance and mass estimates, presenting its results. Section~\ref{sec:discussion} assesses the accuracy of our method and discusses the overall resulting mass and distance distributions, along with limitations and biases. Finally, Section~\ref{sec:conc} provides our conclusions.

\section{Sample \& Data Selection}
\label{sec:dataselect}

Out of the 70 transient black hole sources observed as of December 2023, we selected all sources that underwent at least one typical outburst according to the catalogues of \citet{tetarenko2016}, \citet{corral-santana2016} \footnote{Constantly updated to include latest transient BH-LMXBs}, and \citet{alabarta2021}. This yields an initial sample of 50 potential BH-LMXB sources. For this initial sample of targets, we created separate light curves of every catalogued outburst using available archival data from \maxi/GSC, \swift/XRT and/or \rxte/PCA. The light curve creation and analysis are described below (Sections \ref{sec:lightcurvecreate} and \ref{sec:state_identify}). Using the light curve analysis, we could further limit our sample to sources with an identifiable soft state in at least one of the observed outbursts. 

Moreover, we excluded sources from our sample whose spectra were later found to be of low quality for our analysis. Namely, sources that had less than 10 data points after binning to a signal-to-noise ratio of 3 or sources that were unable to obtain an X-ray spectral description with $\chi_{\text{red}}^2<1.4$ using our models. After applying these filtration criteria, we were left with a final sample of 26 unique sources for our analysis, with some sources having multiple outbursts or the same outburst detected in multiple instruments.

As we will see in Section~\ref{sec:method-overall}, our distance estimation method is ideally based on the spectral modelling of the soft and the soft-to-hard transition state spectra during a typical outburst from the same instrument. However, a lower-quality distance estimate can be obtained from the soft state alone. In our final sample of 26 sources, the 18 sources with both soft state and transient state spectral information can be found in Table~\ref{tab:table1_result}, and the 8 sources with only soft spectral information are listed in Table~\ref{tab:table2_result}.

We have chosen to consider data from \maxi/GSC, \swift/XRT and \rxte/PCA for our study as these instruments have observed the majority of BH-LMXB outbursts. Additionally, \maxi/GSC and \swift/XRT continue to make new discoveries. We did not simultaneously model data or combine estimates coming from these different instruments for the same source/outburst. This allows us to assess any biases arising from using a specific detector since we have used data of the same source/outburst from multiple detectors whenever possible.

\section{Data Reduction \& Preparation}\label{sec:datared}
This section describes how the \maxi/GSC, \swift/XRT, and \rxte/PCA data for this study have been reduced and how light curves and spectra of certain points in the hardness intensity diagram were created. 

All locally downloaded data were reduced and analyzed using HEASoft v6.32.1 \citep{heasoft} with the latest CALDB files (as of Dec 2023) for each of the used instruments.

\subsection{Light curve creation}\label{sec:lightcurvecreate}
\subsubsection{\maxi/GSC}
For \maxi/GSC, data were obtained using the \texttt{mxdownload\_wget} command and reduced using the \texttt{mxproduct} pipeline both provided in HEASoft from the \maxi\, team\footnote{\url{https://www.darts.isas.jaxa.jp/astro/maxi/analysis/dataanalysis.html}}. Source and background regions were determined using the {\it MAXI} on-demand \footnote{\url{http://maxi.riken.jp/mxondem/}} web tool \citep{nakahira2012}. For each source, daily average spectra were then produced using the tool. Daily light curves in two bands, 2-4 keV and 4-20 keV, were created from the reduced daily spectra. 

\subsubsection{\swift/XRT}
For those outbursts covered by \swift/XRT, light curves were obtained using the Swift/XRT web tool \citep{xrt-lc1,xrt-lc2}, which used HEASoft v6.29 at the time we produced the light curves. For each outburst, light curves in the energy range of 0.3-4 keV and 4-10 keV were produced with a binning by observation. 

\subsubsection{\rxte/PCA}
Ready-analyzed \rxte/PCA light curves were downloaded using the HEAVENS\footnote{\url{https://www.isdc.unige.ch/integral/heavens}} tool, which is part of the INTEGRAL Science Data Centre \citep[ISDC;][]{rxte_data1,rxte_data2}.
The 2-10 keV and 10-20 keV light curves were binned in 1-day bins and limited to counts from the Proportional Counter Unit (PCU) 2 detector and the top layer.

\subsection{Spectral State Identification}\label{sec:state_identify}
After obtaining the light curves for each instrument, we calculate the hardness ratio (HR) with time for individual outbursts by dividing the number of counts in the harder band light curve by the softer band light curve counts in a given time bin. The soft and transition periods are then identified manually using the values of the daily HR. In particular, following the creation of the HR for each outburst and instrument, we first identify the entirety of the soft state period, the time when there is a consistently lower HR relative to the HR at the start of the outburst. The soft-to-hard transition period is then determined by identifying the first few days when the HR recovers from the consistently low values. This selection is performed by visual inspection of the HR plot.

In order to validate this manual identification method, we also ran a K-means clustering algorithm where we use the HR and intensity (total flux) as input features similar to a recent study by \citet{sreehari2021}. We use a total of 4 clusters corresponding to the hard state, hard-to-soft intermediate state, soft state, and the soft-to-hard intermediate state. We then let the algorithm assign each data point to one of these clusters. For most instances, there was a consensus between the manual identification process and the results derived from K-means clustering. However, there were a few cases where the manual identification did not align with the K-means clustering outcome. In these situations, we relied on the K-means clustering results to identify the appropriate periods of the soft state and the soft-to-hard transition. 

Once the soft state and ideally the transition state dates are identified (Table~\ref{tab:fit_table}), we then produce corresponding spectra for further analysis as described in the next sub-section.  

\subsection{X-ray Spectra creation}
\subsubsection{\maxi/GSC}
For {\it MAXI}, we have already produced daily spectra to obtain a light curve, and we utilize those for our soft and transition state spectra. 

However, most of the individual daily spectra did not have the necessary statistics to provide meaningful spectral parameters. Thus, spectra with similar hardness ratios (Photons s$^{-1}$ m$^{-2}$ of 4-20 keV to 2-4 keV) were grouped together such that the combined spectrum has at least 2000 net (background-subtracted) photons. Similarly, we required a minimum of 2000 net counts for the soft-to-hard transition period. The final combined spectra were then binned to a signal-to-noise (S/N) ratio of 3.

\subsubsection{\swift/XRT}
For \swift/XRT, we downloaded the Windowed Timing (WT) mode observations matching the soft and transition state times and reduced the data locally using the \texttt{xrtpipeline}. To create a spectrum, the source region for non-piled-up observations was taken to be a circle with a radius $r_s$ of 50 pixels centered on the source coordinates. While the background region was taken to be an annulus with an inner radius $r_{b,\text{in}}$ of 80 pixels and an outer radius $r_{b,\text{out}}$ of 120 pixels. Following the guidelines by the \swift~team\footnote{\url{https://www.swift.ac.uk/analysis/xrt/backscal.php}}, both the source and background had their \texttt{BACKSCAL} keyword edited to $2r_s$ and $r_{b,\text{out}} - r_{b,\text{in}} - 1$, respectively.

For sources with 100 or more counts per second in the resulting spectrum, a pile-up correction was performed similar to method 2 described in \citep{mineo2006} but with a simpler approach. In our approach, we specify an annulus region centered on the source center. The starting inner radius is arbitrarily chosen, while an outer radius is chosen to be 100\,arcseconds for all observations. We then increase or decrease the inner radius by several arcseconds until we have a spectrum with a count rate of less than 100. The background region shape for the piled-up sources is then taken to be the same as the source annulus but sufficiently away from the center. The spectra of the individual observations are much more statistically significant than the individual observations from \maxi, so no single observations were combined. All spectra were binned to an S/N ratio of 3.

\subsubsection{\rxte/PCA}
Data reduction of the observations matching the soft/transition states was done locally by first processing the data into usable Standard2 spectral products using the PCA Standard2 extraction pipeline provided by the \rxte~team\footnote{\url{https://heasarc.gsfc.nasa.gov/docs/xte/recipes2/Overview.html}} and included in HEASoft v6.32.1. We used the latest CALDB file released on May 1, 2023, which automatically adds a systematic error that is recommended by the \rxte~team. The pipeline runs 4 steps: (1) data preparation, where basic filtering and data processing happen, (2) merging observations if using more than one, (3) making a good time filter which excludes bad data points, (4) extracting the final spectral product which is a "Standard2" data product. The Standard2 data products contain the energy and temporal information required for spectral modelling. The spectra were also binned to a S/N ratio of 3 and no observations were combined.

\section{Spectral Analysis}
\label{sec:spec_model}

To ultimately obtain a constraint for the distance to a BH-LMXB from the X-ray spectrum during an outburst. First, we need to model the spectrum during the soft state and, if possible, the soft-to-hard transition.

All spectral analyses presented here were performed in \xspec~\citep[v12.13.1;][]{xspec_ref}. Solar abundance values were set to \textit{wilm} \citep{wilms2000} while the cross-section table was set to \textit{vern} \citep{vern}.  
The built-in Markov Chain Monte Carlo (MCMC) was used to find the uncertainties in all variable spectral parameters. All MCMC chains used the Goodman-Weare (GW) algorithm and had a total length of $2 \times 10^{6}$ with a burn-in of $1\times10^{5}$ and 40 walkers.

In BH-LMXBs, the soft state and the soft-to-hard transition X-ray spectrum in the energy range 0.5-20\,keV can be described well to first order by combining two models: a multi-color disk black body model accounts for the accretion disk emission and a Compton up-scattered emission component \citep{remillard2006}, which a power law can approximate. This intrinsic X-ray spectral model is then further altered by photoelectric absorption along the line of sight. Thus, in \xspec~we implemented this base model as:\\
\texttt{TBabs$\ast$(powerlaw+ezdiskbb)} \textbf{(Model S1)}, \\where the \texttt{TBabs}~\citep{wilms2000} model accounts for the interstellar absorption, a phenomenological \texttt{powerlaw} models the Compton up-scattered photons, and the \texttt{ezdiskbb} \citep{zimmerman2005} models the disk's radiation. In this model set-up, we always fit for the photon index ($\Gamma$), power law normalization, the maximum temperature of the disk ($T_{\text{max}}$; assumed to be $>0.1$ keV), and the disk normalization ($N_{\texttt{ezdiskbb}}$). 

For the \maxi/GSC and \rxte/PCA spectra, the energy ranges used for spectral modelling were 2-20 keV and 3-20 keV, respectively. This leads to the data being unable to constrain the interstellar absorption column, ($N_{\text{H}}$), and the parameter was that kept fixed\footnote{One exception was XTE J1748-2848 in its 1998 outburst (\rxte/PCA data) as the previous $N_{\text{H}}$ ($7.5 \times 10^{22} $ cm$^{-2}$ \citep{miller2001}) was high, and when we fixed it to this value we did not get good fits ($\chi^2 \geq 2$).} to the literature values for each source (see Table~\ref{tab:fit_table} for the used absorption columns). On the other hand, the energy range used for \swift/XRT spectra was 0.7-10 keV, which allowed for $N_{\text{H}}$ to be a free\footnote{An exception to this was MAXI J0637-430 as fitting routine was unable to constrain it in our selected observations. So, we fixed it to $ 0.4 \times 10^{22}$ cm$^{-2}$ \citep{jia2023}.} parameter. 

For the soft state spectra, we expect that the spectrum is very steep, and consequently, $\Gamma$ was only permitted to vary between 1.7 and 3.0. The parameter relevant for the distance estimation from the spectral parameters of this state is the normalization of the disk component. On the other hand, we expect the spectrum to be harder for the transition period, and the photon index was confined to the lower range of 1.0-2.2. The parameter that will be relevant for distance estimation from the transition state spectrum is the 0.5-200 keV power-law flux, as discussed in the next section. 

We note that for some \rxte/PCA sources, the spectrum contained a significant 6.4\,keV iron K line that was affecting the spectral parameters.  Therefore, for those sources, we added a Gaussian component (\texttt{gauss}) centered at 6.4 \,keV and a maximum width of 1\,keV. The full model used for these sources is implemented in \xspec~as:\\ \texttt{TBabs$\ast$(powerlaw+ezdiskbb+gauss)} \textbf{(Model S2)}.

We require the unabsorbed power-law flux of the transition state in the 0.5-200 keV energy band for the distance estimation. The models above were modified to obtain a robust constraint of this flux from the convolution model \texttt{cflux}. We extended the response matrix range using the \texttt{energies} command, so the model calculation covers the entire 0.5-200\,keV band. As the \texttt{cflux} flux value and the power law normalization would be degenerate in the modeling, we adjust the modeling set-up for the transition state. Namely, we freeze the normalization of the unabsorbed model power-law component and apply it as the following: 
\begin{itemize}
\item[] \texttt{TBabs$\ast$(cflux$\ast$powerlaw+ezdiskbb)} \textbf{(Model T1)}

\item[] or, if this does not yield to a satisfactory description,

\item[] \texttt{TBabs$\ast$(cflux$\ast$powerlaw+ezdiskbb+gauss)} \textbf{(Model T2)}. 
\end{itemize}

Using the version of Model 1 or Model 2 corresponding to the state, as described above, we modelled a total of 66 spectra. More specifically, we fitted 39 soft-state spectra and 27 soft-to-hard transition spectra for a total number of 34 unique outbursts. 

Our basic models 1 and 2, were able to describe these spectra well with $\chi^2_{\text{red}}\leq$ 1.4 in all cases and the majority of fits having $\chi^2_{\text{red}}$ very close to 1. A summary of our spectral modeling results can be found in Table~\ref{tab:fit_table} in the appendix. 

From our spectral modeling, we obtain the posterior distribution of the normalization of \texttt{ezdiskbb} ($N_{\texttt{ezdiskbb}}$) by performing MCMC analysis with Gaussian priors based on the best-fit model parameters for each of the soft state spectra. Similarly, for soft-to-hard transition fits, we obtain the posterior distribution of the power law flux in the 0.5-200\,keV range by utilizing the \texttt{cflux} component. For computational efficiency, these resulting distributions were down-sampled from the initial length $2 \times 10^{6}$ (length of the MCMC chain) to $2 \times 10^{4}$ by averaging every 100 adjacent values. The MCMC distributions were then used as inputs for the statistical framework described in Section~\ref{sec:probability_method} to produce distance estimates for each source in our final sample.

\section{Distance Calculation}
In the following, we discuss how the modelling results from the soft and the transition state spectra can be used to obtain a high-quality distance constraint.
\label{sec:method-overall}
\subsection{The Basic Idea}
A fairly simple distance estimation method based on only information from X-ray spectral modelling was used multiple times throughout the literature \citep[see][]{nakahira2012,nakahira2014,oda2019,tominaga2020} to produce a quick constraint on the distance to a BH-LMXB in outburst.

This method is based on the same fundamental idea as many astronomical distance estimation methods: if we measure the flux and know what the intrinsic luminosity should be, we can find the distance. In the soft state, the X-ray spectrum of black hole X-ray binaries can be assumed to be disk-dominated, with only a small portion of the flux potentially originating from the power-law component. Therefore, we can get very accurate information about the total observed flux in this state from the accretion disk flux. 
In practice, the spectral modeling of the soft state spectrum does not return the accretion disk flux directly, but the accretion disk spectrum is modeled and then the disk normalization is used to derive the flux-distance relation. In particular, the normalization of the multi-color disk black body accretion disk model, \texttt{ezdiskbb}, that we have used in our modeling is defined as \citep{zimmerman2005}:
\begin{equation}
N_{\texttt{ezdiskbb}} = \left( \frac{r_{\text{in}} \text{ [km]}}{f_{\text{col}}^2 \left[D/10 \text{ kpc}\right]} \right)^2 \cos i
\label{eq:norm_ezdiskbb}
\end{equation}
where $f_{\text{col}} = 1.7$ (see Section~\ref{sec:limitations} for discussion about this value) is the color (spectral) hardening factor \citep{shimura1995}, $D$ is the distance to the source in kpc, and $i$ is the disk inclination as observed in the sky (e.g. $i=0^\circ$ is face-on). 
As evident from equation~(\ref{eq:norm_ezdiskbb}), the observed disk spectrum depends not only on the distance to the source but also on the inclination and inner radius of the accretion disk. Observational evidence indicates that this inner radius is nearly constant during the soft state and is thought to be equal to the theoretical innermost stable circular orbit \citep[ISCO; e.g.][]{steiner2010}. The location of the ISCO generally depends on the BH mass and spin. However, if we assume a non-rotating BH, then it only depends on mass, where $r_{\text{ISCO}} = 6GM/c^2$, with $M$ being the black hole mass, $G$ is the gravitational constant and $c$ is the speed of light. As a result, the constraint of the normalization of the \texttt{ezdiskbb} model obtained from the spectral modelling can be used to produce a relation between the mass, the inclination, and the distance to the source.

Ideally, we need to constrain the parameter space further to break some of the degeneracy between the distance, the mass, and the inclination for a given spin value. To do this, we make use of an empirical relation first found by \citet{maccarone2003} and later investigated in \citet{dunn2010,kalemci2013,tetarenko2016,vahdat2019}, in which the bolometric luminosity of the state transition from the soft state to the hard state was typically observed to happen at 1-4\,\% of the Eddington luminosity ($L_{\text{Edd}}$). More specifically, \citet{maccarone2003} and recently \citet{vahdat2019} found that the ($\sim 0.5-200$ keV) power law flux in the soft-to-hard spectral state transition has been consistently observed to have a mean of $~2\%$ of the Eddington luminosity. A similar study by \citet{tetarenko2016}, where they considered not only the power law flux but also the emission of the disk, also found a small, mean percentage of $\sim$3\%. From modeling the X-ray spectrum at the transition time, we are able to obtain the flux. Since the flux is related to the luminosity via the squared distance and the Eddington luminosity depends on mass, we obtain an additional relation between distance and mass. For instance, using the (1-4\,\%) range found by \citet{maccarone2003}:
\begin{equation}
    4 \pi D^2 F_{\text{trans}} = f_{\text{Edd}} \times 1.3 \times 10^{38} \frac{M}{M_{\odot}} \text{erg/s}
     \label{eq:trans_relation}
\end{equation}
where $F_{\text{trans}}$ is the bolometric flux at the soft-to-hard transition, $f_{\text{Edd}}$ is the Eddington fraction which is 0.01-0.04, and we have used $L_{\text{Edd}} = 1.3 \times 10^{38} \frac{M}{M_{\odot}}$ erg/s and $L = 4 \pi D^2 F$ (isotropic emission).

So theoretically, one can obtain a constraint on the distance from the spectral modeling results and equations \ref{eq:norm_ezdiskbb} and/or \ref{eq:trans_relation} after making assumptions for $i$ and $M$. However, this simple approach assumes a non-rotating black hole and ignores the specific statistical distribution of the spectral parameters and other relativistic effects, as discussed below. 

\subsection{Limb Darkening and Relativistic Corrections}\label{sec:method-correc}
In the previous section, limb-darkening or relativistic effects were not considered, yet these are important effects that influence the accuracy of this type of distance estimate. Despite their importance, applications of the method outlined in the previous section have mostly ignored these effects. In this study, we take these corrections into account by following the work done by \citet{salvesen2021} \citep[see also][]{zhang1997} and re-writing the normalization of \texttt{ezdiskbb} as:
\begin{equation}
    N_{\texttt{ezdiskbb}} = \left( \frac{r_{\text{in}} \text{ [km]}}{f_{\text{col}}^2 \left[D/10 \text{ kpc}\right]} \right)^2 \cos i~\Upsilon(i)~g_{GR}(a,i)~g_{NT}(a)
    \label{eq:norm_ezdiskbb_corrections}
\end{equation}
where $\Upsilon(i)= \dfrac{1}{2} + \dfrac{3}{4} \cos i$ is the limb darkening law assuming a disk atmosphere that is plane-parallel, semi-infinite and electron scattered \citep{chandrasekhar1960}, $g_{GR}(a,i)$ is a correction factor that takes into account the relativistic effects on photon propagation, and $g_{NT}(a)$ is a correction factor that takes into account the relativistic radial structure of the accretion disk. We note that $g_{GR}(a,i)$ depends on the black hole spin, $a$, and the disk inclination, $i$, while $g_{NT}(a)$ only depends on the black hole spin. These dependencies allow us to pre-calculate these correction factors for possible values of spin and inclination regardless of the properties of any particular source. In fact, these corrections were tabulated by \citet{salvesen2021} and made available in a public repository\footnote{\url{https://github.com/gregsalvesen/bhspinf/tree/main/data/GR}}. In their work, the authors utilize the \texttt{kerrbb} model \citep{kerrbb} and numerical integration to calculate the flux, taking into account the relativistic effects. They then determine the correction factors by calculating the ratio of fluxes with and without these effects. In our work, we use their tabulated correction values to account for the relativistic effects.

\subsection{Obtaining Distance/Mass Probability Distributions}
\label{sec:probability_method}
Rather than just making assumptions using a few discrete values for $a$, $i$, $M$, we use an approach that considers all of the uncertainties entering from the calculations and yields a probability distribution of the distance values for each source. 

It starts with the MCMC spectral modeling error analysis of the soft state spectrum of a source, which produces a probability distribution of $N_{\texttt{ezdiskbb}}$. We then obtain a joint mass/distance probability density function  $p_{M,D}(M,D|\text{soft})$ by calculating the distance using equation~(\ref{eq:norm_ezdiskbb_corrections}) over sampled $i$, $M$, and $a$ values. Similarly, the spectral modelling followed by an MCMC error analysis of the source when it goes through the soft-to-hard transition produces a probability distribution of $F_{\text{trans}}$. Once again, using an assumed distribution of black hole mass, as well as an Eddington fraction $f_{\text{Edd}}$, we find another mass/distance probability density function $p_{M,D}(M,D|\text{trans})$ using equation~(\ref{eq:trans_relation}). Finally, we assume that the best estimate for the true mass/distance joint probability distribution is:
\begin{equation}
    p_{M,D}(M,D) = K p_{M,D}(M,D|\text{soft}) p_{M,D}(M,D|\text{trans})
    \label{eqn:combined_soft_trans_pdf}
\end{equation}
where $K$ is some normalization factor that makes the combined probability density function $p_{M,D}(M,D)$ integrated over all possible distance and mass values equal to one. 

\begin{figure}
	\includegraphics[width=\columnwidth]{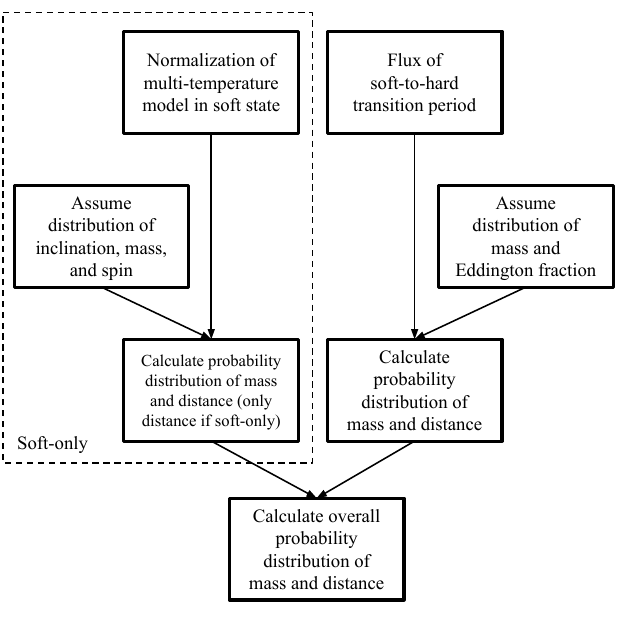}
    \caption{This diagram summarizes the statistical framework, which produces mass/distance probability distributions. The framework is discussed in detail in section \ref{sec:probability_method}.} 
    \label{fig:probability_chart1}
\end{figure}

We can find the marginal probability distribution of distance or mass by integrating over all possible values of the other variable. For example, to obtain the marginal probability density for the distance we integrate over all possible masses:
\begin{equation}
    p_D(D) = \int p_{M,D}(M,D)~dM
\end{equation}

Furthermore, equation~(\ref{eqn:combined_soft_trans_pdf}) shows that unless $p_{M,D}(M,D|\text{soft})$ and $p_{M,D}(M,D|\text{trans})$ are both non-zero at all distances $D$, we in fact gain new information on the mass probability density function. So, we obtain a refined mass distribution by requiring that the only valid distance values of a certain source are where both constraints have non-zero probability density values. This approach is analogous to Bayesian inference tools, where we obtain a posterior distribution based on the prior guess.

If only a constraint from the soft state spectrum exists, we follow the same procedure until we obtain a mass and distance probability distribution. As we do not have a second mass and distance probability, the distribution of the combination step does not take place. An overview diagram of the process is shown in Fig.~\ref{fig:probability_chart1}.

As mentioned before, we need to use probability distributions for the inclination, black hole spin, Eddington fraction, and black hole mass. For the inclination, we assume an isotropic distribution in the range (0\textdegree, 84\textdegree) where the probability of observing a given inclination is given by $p_i(i) = \sin i$. We chose this specific inclination range since it is very unlikely to observe the disk spectrum at very high inclinations (it will be very faint). Additionally, the correction values $g_{GR}(a,i)$ and $g_{NT}(a)$ in equation~(\ref{eq:norm_ezdiskbb_corrections}) are calculated by utilizing the \texttt{kerrbb} model in \xspec~which only allows for angles 0\textdegree~to 85\textdegree. We note here that this inclination prior could be improved in two ways: universally, by accounting for observational selection effects (a topic beyond the present paper's scope) or, specifically, for individual sources, by considering a probable inclination distribution for the specific source. Notably, we draw attention to a method proposed by \citet{munoz-darias2013} in which the shape of the HID track can be used as an estimator for inclination. This method is of particular relevance given that both relativistic effects and limb-darkening significantly alter the observed pattern. In future work, we aim to conduct a detailed systematic study where we use these HID properties and develop them into a method yielding inclination constraints.

For the black hole spin parameter, we assume a uniform distribution in the range 0.0 to 0.998, where all the spin values in that range are equally likely: $p_a(a) = 1/(0.998-0)$. This range of spins was chosen because of the rarity of BH-LMXBs that are estimated to have retrograde spin \citep{reynolds2021}. However, we investigate the impact of allowing for retrograde spin on our results in Section~\ref{sec:limitations}.

For the Eddington fraction at the transition point, we assume a uniform distribution in the range 0.01 to 0.04 according to \citet{maccarone2003}: $p_{f_{\text{Edd}}}(f_{\text{Edd}}) = 1/(0.04-0.01)$. We chose to use a range of 1-4\% as it captures most of the observed variations in the transition luminosities. We also note that we use only the power law flux following both \citet{maccarone2003} and \citet{vahdat2019}.  

Finally, for the black hole mass distribution, we assume a Gaussian prior centered at 7.8 \ms with a standard deviation of 1.2 \ms. This particular Gaussian distribution was selected based on the work of \citet{ozel2010}, \citet{farr2011}, and \citet{kriedberg2012}. 

In all our calculations of $p_{M,D}(M,D)$, we use these distributions to sample 101, 100, 101, 1000 values for the inclination, spin, Eddington fraction, and black hole mass, respectively.

\begin{figure*}
 \includegraphics{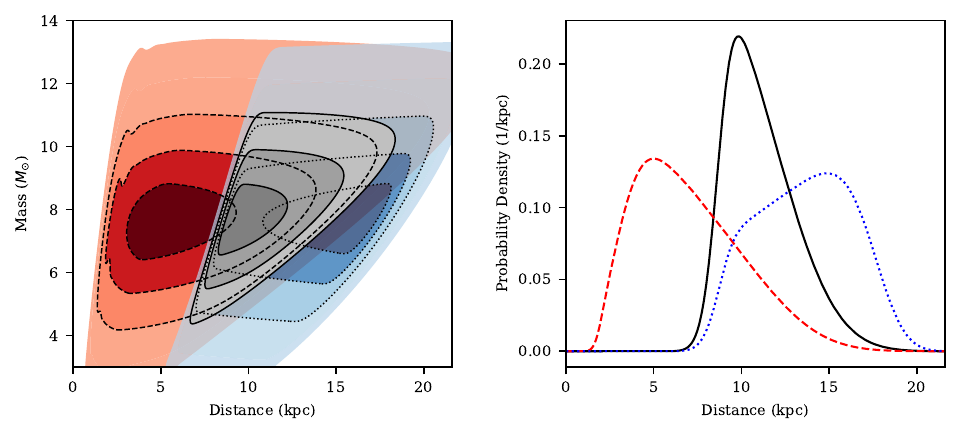}
    \caption{This figure shows the probability distributions resulting from applying our method to the \rxte/PCA observations of Swift J1539.2-6227 during its 2008/2009 outburst. \textbf{Left:} The joint mass/distance probability distribution. The red region corresponds to the probability distribution from only the soft state constraint, while the blue region shows the probability distribution from only the transition luminosity. The grey region visualizes the combined probability distribution that results when both distributions are assumed to be true, equation (\ref{eqn:combined_soft_trans_pdf}). The contour levels shown are 1$\sigma$, 2$\sigma$, and 3$\sigma$.
    \textbf{Right:} The marginalized distance probability distribution. The red line corresponds to the probability distribution from only the soft state constraint, while the blue line indicates the probability distribution from only the transition luminosity. The black line is the probability distribution that results by combining both constraints.}
    \label{fig:probability_example}
\end{figure*}

As shown in the example in Fig.~\ref{fig:probability_example}, we plot the marginal probability distributions of distance from both constraints (red/dashed for soft, blue/dotted for transition) and then obtain the best estimate of the marginal probability by multiplying both distributions (black/solid). Similarly, for the joint mass/distance distribution, we plot the 2D contours of the soft constraint (red/dashed contours) and the transition constraint (blue/dotted contours). We then multiply both of the joint probabilities to obtain the best estimate of the joint mass/distance probability distribution (black/solid contours).

\subsubsection{Sources with Multiple Outbursts}
\label{sec:method-multioutbursts}
Some sources in our sample have been observed during multiple outbursts. To obtain the best estimate for the true mass/distance joint probability distribution, we first find mass/distance probability distribution, $p^{(j)}_{M,D}(M,D)$, for each outburst $j$ according to equation~(\ref{eqn:combined_soft_trans_pdf}), we then find the overall mass/distance probability distribution by multiplying them and normalizing again:
\begin{equation}
    p_{M,D}(M,D) = \Tilde{K} \prod^{n}_{j=1} p^{(j)}_{M,D}(M,D)
    \label{eqn:combined_outbursts_pdf}
\end{equation}
where $n$ is the total number of outbursts and $\Tilde{K}$ is another normalization factor. To produce the overall estimate of only the soft state or the soft-to-hard transition mass-distance joint probability distribution, we use equation~(\ref{eqn:combined_outbursts_pdf}) but with only the soft state or the soft-to-hard transition probabilities for each outburst (see for example the joint probability distribution for GX 334-9 in Figs.~\ref{fig:per_sources_distance1}, and~\ref{fig:per_sources_mass_distance1}).

\subsection{Results}
\label{sec:results}
\begin{table*}
\caption{Summary of distance and mass estimates when using both the soft state and the soft-to-hard transition spectral information. For each source, we show the median distance/mass along with their 1$\sigma$ statistical uncertainties (see Section~\ref{sec:limitations} for a discussion of systematic errors). The probability distributions are shown in Figs.~\ref{fig:per_sources_distance1}, and~\ref{fig:per_sources_mass_distance1}.}
\label{tab:table1_result}
\begin{filecontents*}{tables/Table1.csv}
Name,Outburst(s) Year,med-dist-maxi,lower-err-maxi,upper-err-maxi,med-dist-xrt,lower-err-xrt,upper-err-xrt,med-dist-rxte,lower-err-rxte,upper-err-rxte,med-mass-maxi,mass-lower-err-maxi,mass-upper-err-maxi,med-mass-xrt,mass-lower-err-xrt,mass-upper-err-xrt,med-mass-rxte,mass-lower-err-rxte,mass-upper-err-rxte
GRS 1739-278,2014,,,,12.74,2.42,3.39,,,,,,,7.92,1.16,1.16,,,
GX 339-4,2004-2005,,,,,,,7.61,0.6,0.77,,,,,,,8.73,0.95,1.02
,2007,,,,7.29,1.24,1.8,,,,,,,7.97,1.16,1.16,,,
,2010-2011,7.51,0.86,1.19,,,,8.54,0.48,0.6,8.24,1.09,1.12,,,,9.55,0.77,0.87
,2015-2016,6.39,0.82,1.09,6.45,1.3,1.72,,,,8.17,1.11,1.14,7.87,1.14,1.17,,,
,2019-2020,7.72,1.9,1.93,6.76,1.18,1.58,,,,7.99,1.16,1.17,8,1.14,1.17,,,
,Combined,7.04,0.54,0.65,6.61,0.72,1.05,8.13,0.3,0.34,8.04,0.65,0.66,7.78,0.68,0.68,9.06,0.47,0.51
H 1743-322,2013,9.14,2.22,2.22,,,,,,,7.53,1.19,1.21,,,,,,
IGR J17091-3624,2011-2013,,,,11.81,2.25,2.43,,,,,,,7.34,1.21,1.21,,,
,2016,,,,17.92,3.93,3.21,,,,,,,7.46,1.21,1.21,,,
,Combined,,,,14.16,1.94,2.03,,,,,,,7.36,0.86,0.86,,,
MAXI J0637-430,2019,,,,9.91,1.1,1.57,,,,,,,8.21,1.11,1.12,,,
MAXI J1305-704,2012,,,,10.54,2.42,2.35,,,,,,,7.65,1.19,1.19,,,
MAXI J1348-630,2019,2.42,0.16,0.2,,,,,,,9.26,0.95,0.99,,,,,,
MAXI J1659-152,2010-2011,,,,,,,6.73,1.46,1.19,,,,,,,7.51,1.19,1.21
MAXI J1820+070,2018,2.52,0.4,0.49,2.64,0.56,0.71,,,,8.31,1.12,1.14,7.82,1.16,1.17,,,
MAXI J1828-249,2013,6.44,1.41,1.99,,,,,,,7.85,1.17,1.17,,,,,,
Swift J1539.2-6227,2008-2009,,,,,,,10.8,1.62,2.4,,,,,,,8.02,1.16,1.16
Swift J1842.5-1124,2008-2009,,,,,,,8.12,1.67,2.2,,,,,,,7.85,1.16,1.17
Swift J1910.2-0546,2012-2013,3.8,0.23,0.25,,,,,,,10.05,0.88,0.94,,,,,,
XTE J1650-500,2001-2002,,,,,,,4.38,0.26,0.34,,,,,,,9.25,0.26,0.34
XTE J1652-453,2009,,,,,,,6.69,0.43,0.54,,,,,,,9.18,0.85,0.95
XTE J1748-288,1998,,,,,,,8.66,1.98,1.84,,,,,,,7.63,1.21,1.19
XTE J1752-223,2009-2010,,,,,,,7.11,0.25,0.27,,,,,,,12,0.68,0.77
XTE J1908+094,2013-2014,,,,11.52,2.74,1.93,,,,,,,8.45,1.14,1.12,,,
\end{filecontents*}
\pgfplotstabletypeset[col sep=comma,
columns/Name/.style={string type,column name = Source,column type=l},columns/Outburst(s) Year/.style={string type,column name=Outburst Year(s)$^{a}$}, 
every head row/.style={before row=\toprule,after row=\midrule},
every last row/.style={after row=\bottomrule},
create on use/distance-maxi/.style={
        create col/assign/.code={
            \edef\entry{\thisrow{med-dist-maxi}$_{-\thisrow{lower-err-maxi}}^{+\thisrow{upper-err-maxi}}$}
            \pgfkeyslet{/pgfplots/table/create col/next content}\entry}},
        columns/distance-maxi/.style={string type,column name=\maxi,string replace={$_{-}^{+}$}{--},string replace={none$_{-none}^{+none}$}{--},string replace={$_{-0.00}^{+0.00}$}{--}
        },
create on use/distance-xrt/.style={
        create col/assign/.code={
            \edef\entry{\thisrow{med-dist-xrt}$_{-\thisrow{lower-err-xrt}}^{+\thisrow{upper-err-xrt}}$}
            \pgfkeyslet{/pgfplots/table/create col/next content}\entry}},
        columns/distance-xrt/.style={string type,column name=\swift,string replace={$_{-}^{+}$}{--},string replace={none$_{-none}^{+none}$}{--},string replace={$_{-0.00}^{+0.00}$}{--}
        },
create on use/distance-rxte/.style={
        create col/assign/.code={
            \edef\entry{\thisrow{med-dist-rxte}$_{-\thisrow{lower-err-rxte}}^{+\thisrow{upper-err-rxte}}$}
            \pgfkeyslet{/pgfplots/table/create col/next content}\entry}},
        columns/distance-rxte/.style={string type,column name=\rxte,string replace={$_{-}^{+}$}{--},string replace={none$_{-none}^{+none}$}{--},string replace={$_{-0.00}^{+0.00}$}{--}
        },
,create on use/mass-maxi/.style={
        create col/assign/.code={
            \edef\entry{\thisrow{med-mass-maxi}$_{-\thisrow{mass-lower-err-maxi}}^{+\thisrow{mass-upper-err-maxi}}$}
            \pgfkeyslet{/pgfplots/table/create col/next content}\entry}},
        columns/mass-maxi/.style={string type,column name=\maxi,string replace={$_{-}^{+}$}{--},string replace={none$_{-none}^{+none}$}{--},string replace={$_{-0.00}^{+0.00}$}{--}
        },
create on use/mass-xrt/.style={
        create col/assign/.code={
            \edef\entry{\thisrow{med-mass-xrt}$_{-\thisrow{mass-lower-err-xrt}}^{+\thisrow{mass-upper-err-xrt}}$}
            \pgfkeyslet{/pgfplots/table/create col/next content}\entry}},
        columns/mass-xrt/.style={string type,column name=\swift,string replace={$_{-}^{+}$}{--},string replace={none$_{-none}^{+none}$}{--},string replace={$_{-0.00}^{+0.00}$}{--}
        },
create on use/mass-rxte/.style={
        create col/assign/.code={
            \edef\entry{\thisrow{med-mass-rxte}$_{-\thisrow{mass-lower-err-rxte}}^{+\thisrow{mass-upper-err-rxte}}$}
            \pgfkeyslet{/pgfplots/table/create col/next content}\entry}},
        columns/mass-rxte/.style={string type,column name=\rxte,string replace={$_{-}^{+}$}{--},string replace={none$_{-none}^{+none}$}{--},string replace={$_{-0.00}^{+0.00}$}{--}
        }, 
every head row/.style={
        before row={
            \toprule
            & & \multicolumn{3}{c}{Estimated D (kpc)$^{b}$} & \multicolumn{3}{c}{Estimated Mass (\ms)}\\
        },
        after row=\midrule,
    },
    every last row/.style={
        after row=\bottomrule},after row ={[0.6ex]}
,columns={Name,Outburst(s) Year,distance-maxi,distance-xrt,distance-rxte,mass-maxi,mass-xrt,mass-rxte}
]{tables/Table1.csv}
\begin{tablenotes}
\small 
\item \textbf{Notes.} $^{a}$ "Combined" means the resulting distance or mass estimate is obtained though the combination of probability distributions from all the outbursts using equation~\ref{eqn:combined_outbursts_pdf}.\\
$^{b}$ Quoted errors are statistical-only.

\end{tablenotes}
\end{table*}
\begin{table}
\setlength{\tabcolsep}{3pt}
\caption{Summary of distance estimates using the soft state spectral information only for all sources used in this work. For each source, we show the median distance along with 1$\sigma$ errors statistical uncertainties (see Section~\ref{sec:limitations} for a discussion of systematic errors). The probability distribution of distance are shown in Fig.~\ref{fig:per_sources_distance_soft}.}
\label{tab:table2_result}
\begin{filecontents*}{tables/Table2.csv}
Name,Outburst(s) Year,soft-med-dist-maxi,soft-lower-err-maxi,soft-upper-err-maxi,soft-med-dist-xrt,soft-lower-err-xrt,soft-upper-err-xrt,soft-med-dist-rxte,soft-lower-err-rxte,soft-upper-err-rxte
GRS 1739-278,2014,,,,8.57,3.41,4.77,,,
GX 339-4,2004-2005,,,,,,,3.39,1.35,1.89
,2007,,,,4.7,1.9,2.6,,,
,2010-2011,3.93,1.57,2.22,,,,3.32,1.32,1.84
,2015-2016,3.4,1.37,1.92,4.49,1.78,2.52,,,
,2019-2020,4.76,1.9,2.69,4.29,1.71,2.38,,,
,Combined,3.47,0.95,1.32,3.9,1.06,1.58,2.92,0.79,1.2
H 1743-322,2013,14.75,6.06,8.73,7.86,3.12,4.38,,,
IGR J17091-3624,2011-2013,,,,15.69,4,3,,,
,2016,,,,16.25,3.7,2.64,,,
,Combined,,,,16.97,3.19,2.14,,,
MAXI J0637-430,2019,,,,5.37,2.14,2.98,,,
MAXI J1305-704,2012,,,,11.67,4.64,6.5,,,
MAXI J1348-630,2019,0.89,0.36,0.5,,,,,,
MAXI J1543-564,2011,,,,17,6.77,9.41,15.12,6.01,8.39
MAXI J1659-152,2010-2011,,,,6.74,2.68,3.74,11.08,4.42,6.19
MAXI J1727-203,2018,1.47,0.61,0.9,,,,,,
MAXI J1803-298,2021,7.19,2.9,4.13,,,,,,
MAXI J1820+070,2018,1.28,0.51,0.72,1.92,0.76,1.07,,,
MAXI J1828-249,2013,4.34,1.86,2.83,,,,,,
Swift J1539.2-6227,2008-2009,,,,,,,6.61,2.63,3.69
Swift J1727.8-1613,2023,1.52,0.61,0.85,1.7,0.68,0.95,,,
Swift J1842.5-1124,2008-2009,,,,,,,6.13,2.44,3.42
Swift J1910.2-0546,2012-2013,1.17,0.47,0.66,,,,,,
XTE J1650-500,2001-2002,,,,,,,1.78,0.7,0.99
XTE J1652-453,2009,,,,,,,2.73,1.09,1.53
XTE J1720-318,2003,,,,,,,2.98,1.19,1.65
XTE J1748-288,1998,,,,,,,10.53,4.11,5.75
XTE J1752-223,2009-2010,1.16,0.65,0.93,,,,1.87,0.75,1.04
XTE J1817-330,2006,,,,,,,3.25,1.3,1.8
XTE J1818-245,2005,,,,,,,2.21,0.9,1.29
XTE J1859+226,1999-2000,,,,,,,4.13,1.65,2.29
XTE J1908+094,2013-2014,,,,5.65,2.25,3.15,,,
\end{filecontents*}
\pgfplotstabletypeset[col sep=comma,
columns/Name/.style={string type,column name = Source,column type=l},columns/Outburst(s) Year/.style={string type,column name=Outburst Year(s)$^{a}$}, 
every head row/.style={before row=\toprule,after row=\midrule},
every last row/.style={after row=\bottomrule},
create on use/distance-maxi/.style={
        create col/assign/.code={
            \edef\entry{\thisrow{soft-med-dist-maxi}$_{-\thisrow{soft-lower-err-maxi}}^{+\thisrow{soft-upper-err-maxi}}$}
            \pgfkeyslet{/pgfplots/table/create col/next content}\entry}},
        columns/distance-maxi/.style={string type,column name=\maxi,string replace={$_{-}^{+}$}{},string replace={none$_{-none}^{+none}$}{--},string replace={$_{-0.00}^{+0.00}$}{--},string replace={}{--}
        },
create on use/distance-xrt/.style={
        create col/assign/.code={
            \edef\entry{\thisrow{soft-med-dist-xrt}$_{-\thisrow{soft-lower-err-xrt}}^{+\thisrow{soft-upper-err-xrt}}$}
            \pgfkeyslet{/pgfplots/table/create col/next content}\entry}},
        columns/distance-xrt/.style={string type,column name=\swift,string replace={$_{-}^{+}$}{},string replace={none$_{-none}^{+none}$}{--},string replace={$_{-0.00}^{+0.00}$}{--},string replace={}{--}
        },
create on use/distance-rxte/.style={
        create col/assign/.code={
            \edef\entry{\thisrow{soft-med-dist-rxte}$_{-\thisrow{soft-lower-err-rxte}}^{+\thisrow{soft-upper-err-rxte}}$}
            \pgfkeyslet{/pgfplots/table/create col/next content}\entry}},
        columns/distance-rxte/.style={string type,column name=\rxte,string replace={$_{-}^{+}$}{},string replace={none$_{-none}^{+none}$}{--},string replace={$_{-0.00}^{+0.00}$}{--},string replace={}{--}
        }, 
every head row/.style={
        before row={
            \toprule
            & & \multicolumn{3}{c}{Estimated D (kpc)}\\
        },
        after row=\midrule,
    },
    every last row/.style={
        after row=\bottomrule},after row ={[0.6ex]}
,columns={Name,Outburst(s) Year,distance-maxi,distance-xrt,distance-rxte}
]{tables/Table2.csv}
\begin{tablenotes}[flushleft]
\small 
\item \textbf{Notes.}  $^{a}$ "Combined" means the resulting distance estimate is obtained though the combination of probability distributions from all the outbursts.\\
$^{b}$ Quoted errors are statistical.
\end{tablenotes}
\end{table}

\begin{figure}
	\includegraphics{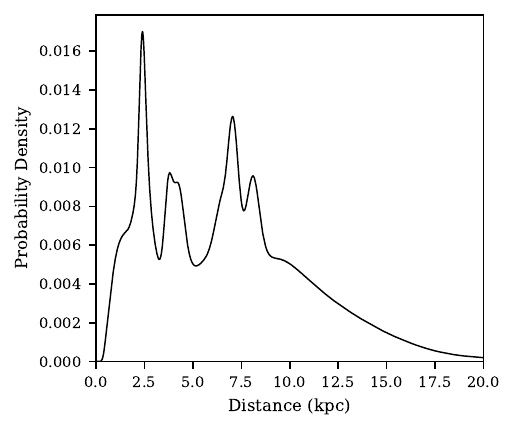}
    \caption{This graph shows the overall heliocentric distance probability distribution, obtained by summing the individual distributions, for the 26 sources in our sample. The plot shows peaks at 2.4, 4.0, 7.1, and 8.1 kpc. We avoid double-counting sources by taking the narrowest probability distribution for 
the sources observed by multiple instruments.}
    \label{fig:overall_dist_1d}
\end{figure}
\begin{figure}
	\includegraphics{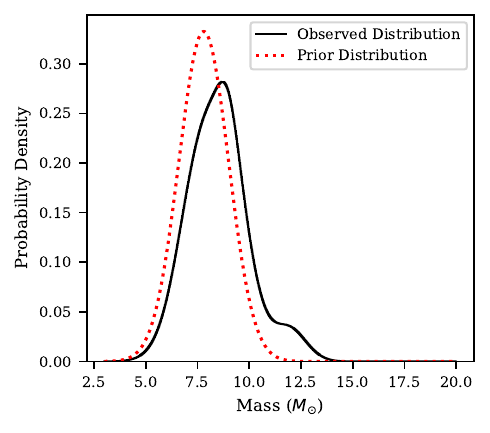}
    \caption{This plot shows the observed overall mass probability distribution (black) and the prior mass distribution (red line). The observed overall distribution shows that the peak shifted from 7.8 to 8.67 \ms~when compared to the peak of the prior. We also see a higher probability density in the range of 10-12.5 \ms~compared to the prior Gaussian distribution N(7.8,1.2).} 
    \label{fig:overall_mass_1d}
\end{figure}

Using the results from our spectral modeling and the method described above, we successfully estimated the distance to 26 sources. Two of the sources, GX 339-4 and IGR J17091-3624, had multiple observed outbursts, so we combined the individual outburst results to an overall distance/mass estimate for these sources as per the method specified in Section~\ref{sec:method-multioutbursts}. 18 of our sources had constraints from both, the soft and the soft-to-hard transition. The distance marginal probability distributions and the joint distance/mass distributions for these 18 sources are presented in Figs.~\ref{fig:probability_example},~\ref{fig:per_sources_distance1}, and \ref{fig:per_sources_mass_distance1} (note that some outbursts was observed by multiple instruments). From these sources, the majority show a large overlap in the distance marginal probability distribution between the constraints from the two states (red line: soft state, blue line: transition point), resulting in somewhat broad combined probability distributions (black line). We note that in most cases, the constraint from the transition flux suggests larger distances than the constraint from the soft state disk normalization. Only for the sources IGR J17091-3624 (\swift/XRT, 2011 and 2016 outbursts) and H 1743-322 (\maxi, 2010 outburst), we find that the soft state disk spectra require distances larger than the transition flux constraints. In particular, for IGR J17091-3624, the soft state constraint suggests a distance larger than 20\,kpc, while the transition flux constraints from the same outbursts require distances in the 12 to 17\,kpc range. The combined distance distribution has a narrow peak around 14 kpc, following the better-constrained transition fluxes. For the case of H 1743-322, the discrepancy is much smaller, with the soft state peak slightly higher than the transition flux peak. 

In addition to analyzing the eighteen sources for which we have constraints from both states, we were also able to find the probability distributions for eight additional sources, using only the soft constraint. In Figure  \ref{fig:per_sources_distance_soft}, we plot the probability distribution of these eight sources. 

Six of these sources (Swift J1727.8-1613, MAXI J1727-203, XTE J1720-318, XTE J1752-223, XTE J1817-330, XTE J1818-245, and XTE J1859+226) appear to have their distance probability distributions peaks in the lower distance range between 1 and 5\,kpc. While the rest show higher distance distributions. We find that the spread of the soft-only distributions is in general wider than the distributions found by combining the probability distributions from both the soft and transition constraints, as one might expect. For example, the narrowest distribution of the source XTE J1752-223 as observed by \maxi~has a 1$\sigma$ interval of (1.02, 2.66) kpc which is considerably larger compared to the narrowest distribution we get for distance using both constraints for the source MAXI J1348-630 also as observed by \maxi~which has a 1$\sigma$ interval of (2.26, 2.62) kpc. As was the case for the two constraint sources, it is evident that the soft state constraint typically necessitates shorter distances.

Additionally, a few of the sources from the 18 that we found both constraints for, had observations from other outbursts/instruments that we were only able to obtain a distance estimate using the soft state. So, we added their results to Fig.~\ref{fig:per_sources_distance_soft}.

A summary of the resulting distance constraints for each source is shown in Tables~\ref{tab:table1_result} and~\ref{tab:table2_result}. Table~\ref{tab:table1_result} shows the results for sources where both the soft state and soft-to-hard transition were utilized, and Table~\ref{tab:table2_result} shows the estimates when using the soft state only. In these tables, we quote the median distance/mass and the 68\% (1$\sigma$) credible interval (statistical errors). The median of the distances using both the soft state and soft-to-hard transition information is 7.47$^{+3.49}_{4.02}$\,kpc, the closest source is MAXI J1348-630 with a distance 2.42$^{+0.2}_{-0.16}$\,kpc and the farthest is IGR J17091-3624 with distance 14.16$^{+2.03}_{-1.94}$\,kpc (combined outbursts). Furthermore, we find that the mass distribution for some of the sources was updated. Using the updated overall mass estimate, we report that the median of all masses is 8.46$^{+1.41}_{1.45}$\,\ms, with lowest mass 7.36$\pm0.86$\,\ms (IGR J17091-3624, combined outbursts) and highest mass 12$^{+0.77}_{-0.68}$\,\ms (XTE J1752-223). 

\begin{figure}
	\includegraphics{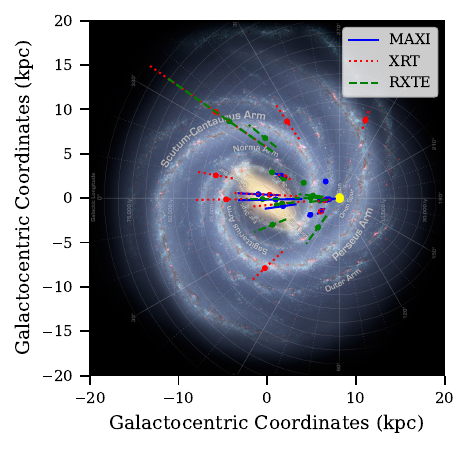}
    \caption{This plot shows the possible location of the 26 sources in our sample projected on the plane of the Milky Way based on their estimated distances and coordinates. Background image: NASA/JPL-Caltech/R. Hurt (SSC/Caltech).} 
    \label{fig:spatial_distribution}
\end{figure} 

Figs.~\ref{fig:overall_dist_1d} and~\ref{fig:overall_mass_1d} show the overall distributions of the distances and masses of all sources in our sample, respectively. Both of these distributions are derived by summing all the individual source probabilities and then re-normalizing such that the total area under curves equals one. Note that for the overall distance distribution (Fig.~\ref{fig:overall_dist_1d}), we have included the soft-only estimates for sources that did not have both constraints. We also avoid double-counting sources by taking the narrowest probability distribution (i.e. estimates with lowest errors) of the common sources across instruments. The resulting distance distribution reveals narrow peaks at 2.4\,kpc and 7.1\,kpc, and slightly wider peaks at 4\,kpc and  8.1\,kpc. We also observe a heavy tail that rabidly decays beyond 15\,kpc. The final mass distribution shows a shift in the observed peak mass value from the prior of 7.8 to 8.7 \ms. Additionally, there is an increase in the probability of finding masses in the 10 to 12.5 \ms~range. 

Furthermore, using the tightest (lowest 1$\sigma$ errors) median distance estimate we obtained for each of the sources in our sample as well as their RA and Dec, we produced the Milky Way spatial distribution of the sources in our sample. We have used the \texttt{astropy.coordinates} method to transform the (ICRS frame) RA, dec, and distance into galactocentric coordinates. We then plotted the possible positions of the sources in the projection of the Milky Way as shown in Fig.~\ref{fig:spatial_distribution}. The plot shows an abundance of sources in the region near or at Bulge. We also see some distant sources observed by \swift/XRT, \rxte/PCA with the majority of sources likely to be associated with the spiral arms.

\section{Discussion}
\label{sec:discussion}

In this study, we have used data from \maxi/GSC, \swift/XRT and \rxte/PCA to develop a robust distance estimation method for black hole low-mass X-ray binaries. Utilizing our distance estimation method and the spectral information of the soft spectral state and the soft-to-hard spectral state transition, we have provided high-quality distance constraints for 26 unique BH-LMXB sources, often exceeding the accuracy of existing constraints (see Section~\ref{sec:improved-dist}). From our results, we constructed a new overall distribution of low-mass black hole X-ray binaries' distances from Earth. Moreover, we have obtained an updated overall black hole mass distribution by combining the distance/mass distribution from both the soft state and soft-to-hard transition state. The \texttt{python} pipeline used to calculate the probability distributions can be found in a public GitHub repository\footnote{\url{https://github.com/ysabdulghani/lmxbd}}. We have also implemented a simplified web version of our improved method for the purpose of providing quick distance estimates for newly discovered BH-LMXBs\footnote{\url{https://solar.physics.montana.edu/youssef/lmxbd/}}.

\subsection{Method Accuracy}
\subsubsection{Comparison to Previous Estimates}\label{sec:comp_to_prev}
\begin{figure*}
 \includegraphics{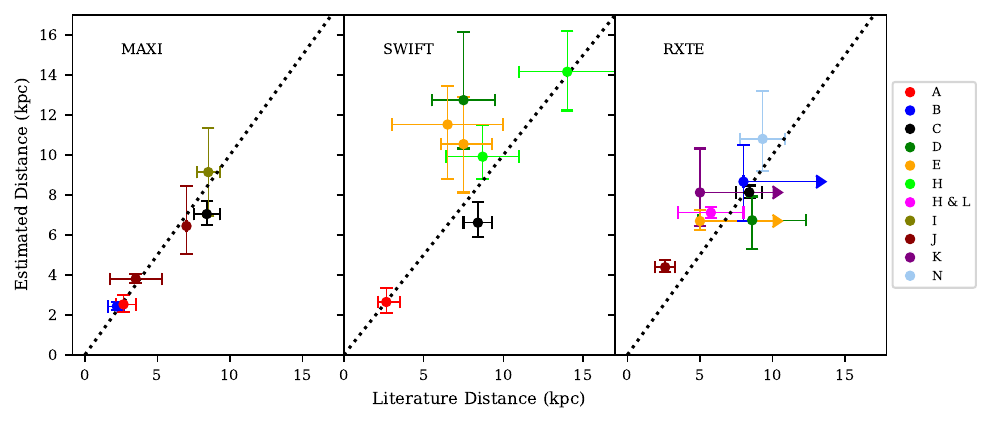}
    \caption{This plot compares the distance constraints obtained from \maxi/GSC, \swift/XRT, and \rxte/PCA data (this work) and distance estimates found in the literature. The dashed line is $x=y$. The error bars in the y-axis correspond to 1$\sigma$ while the error bars in the x-axis correspond to the quoted errors in each of the respective literature works used, where arrows indicate either lower or upper limits. References for the literature distance can be found in Table~\ref{tab:comp_table}. The colors correspond to the method(s) used to estimate the literature distance. \textbf{Methods.} A: Gaia parallax, B: H\,{\sc i} absorption, C: Reflection \& continuum-fitting, D: Interstellar extinction, E: Companion star, F: X-ray decay time scale, G: X-ray luminosity \& disk temperature, H: Hard-to-soft transition luminosity, I: Proper Motion (Jet), J: Soft state and soft-to-hard transition without corrections, K: Radio \& X-ray correlation, L: Assumed to be at Galactic Bulge. M: Flux Scaling, O: Soft state \& peak luminosity.}
    \label{fig:estimate_vs_lit}
\end{figure*}
\begin{figure*}
 \includegraphics{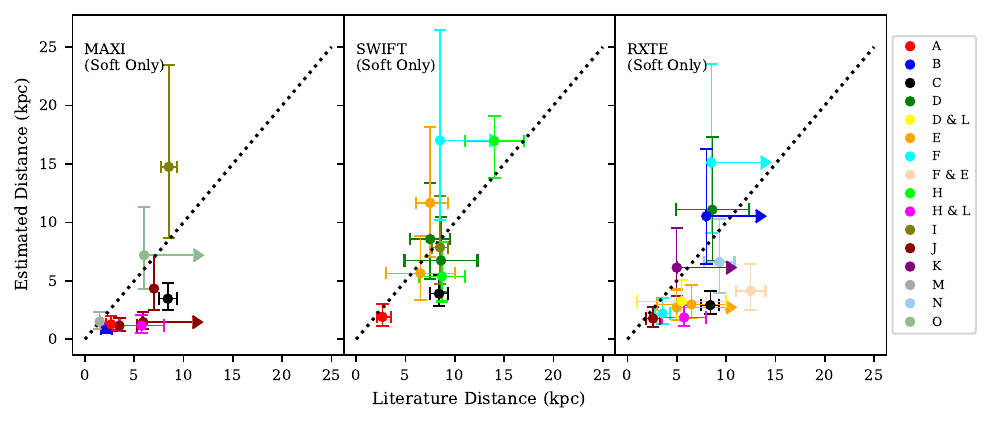}
    \caption{Same as Fig.~\ref{fig:estimate_vs_lit} but for sources/outbursts, where only the soft constraint was found. References for the literature distance are in Table~\ref{tab:comp_table}. The colors correspond to the method(s) used to estimate the literature distance. \textbf{Methods.} A: Gaia parallax, B: H\,{\sc i} absorption, C: Reflection \& continuum-fitting, D: Interstellar extinction, E: Companion star, F: X-ray decay time scale, G: X-ray luminosity \& disk temperature, H: Hard-to-soft transition luminosity, I: Proper Motion (Jet), J: Soft state and soft-to-hard transition without corrections, K: Radio \& X-ray correlation, L: Assumed to be at Galactic Bulge, M: Flux Scaling, O: Soft state \& peak luminosity.}
    \label{fig:estimate_vs_lit_soft}
\end{figure*}

First, we want to assess the agreement of our newly determined distance estimates with existing distance constraints. To facilitate an easy comparison, 
we plot the estimated distance from each instrument against the most accurate literature distance in Fig.~\ref{fig:estimate_vs_lit}. We limited ourselves to those sources, where we were able to use both spectral states in our analysis.

From the graph, it is evident that distances derived from \maxi~and \rxte~data generally show good agreement with previous estimates. They are either consistent within errors or almost equal within the error bars. Most notably, our estimates for MAXI J1820+070 (red point) and MAXI J1348--630 (blue point) using \maxi/GSC data are in perfect agreement with the highly accurate Gaia parallax distance \citep{gaiaed3,tetarenko2021} and H\,{\sc i} absorption distance \citep{chauhan2021a}, respectively.  

Distances derived from \swift~data also show good agreement for three sources, which include MAXI J1820+070. On the other hand, two sources at distances $\sim 7$\,kpc appear to have considerably higher distance estimates from our method than the previous measurements. These two sources are XTE J1908+094 (2013 outburst, orange point) and GRS 1739--278 (2014 outburst, green point). Below, we take a closer look at these sources and their previous distance constraints.

XTE J1908+094 was first discovered in February 2002 \citep{woods2002}. The source later went into two other outbursts in 2013 and more recently in 2019 \citep{krimm2013, rodriguez2019}. The first attempt to estimate the distance to this source was made soon after its discovery in 2002, where \citet{zand2002} used the outburst peak luminosity to derive a lower limit of 3\,kpc. Later, in 2006, \citet{chaty2006} conducted a detailed astrometric analysis and identified a likely candidate for the companion star, which they called "object 1". Using a $J-K'$, $M_{K'}$, absolute colour-magnitude diagram, they determined that "object 1" is an intermediate or late-type main-sequence star. With spectral type between A and K. Consequently, they determined that its distance should be in the range of 3-10\,kpc. This estimate is what we compare to our result. Using our method, we measure a distance of 11.52$^{+1.93}_{-2.74}$ kpc. Although we estimate a median distance considerably higher than the average of their lower and upper limits, the two estimates are still in agreement within the error bars. Moreover, we point to the fact that according to \citet{jonker2004}, using template non-rotating stars to determine the spectral type could lead to underestimation of distance. So if the distance was underestimated by \citet{chaty2006}, that makes our estimates consistent. It is also worth noting that using the same soft-to-hard transition luminosity empirical relation by \citet{maccarone2003} that we used, \citet{curran2015} found an almost identical range (7.8 - 13.6 kpc) to what we found. However, our transition state probability distribution of XTE J1908+094, which is shown in Fig.~\ref{fig:per_sources_distance1}, indicates that the possible distances using only the transition state flux are significantly higher than they found. This is evident from the broad peak of the distribution that appears to be in the range of 11-24 kpc. This discrepancy with the estimation of \citet{curran2015} is most likely due to our different determination of the state transition date, which led them to use a higher transition flux than we used. In their opinion, the state transition was not observed by \swift/XRT due to its proximity to the sun at that period and that in the earliest observation after this pause, the source had already returned to its hard state. In contrast, our HR analysis determined that the source was still transitioning into the hard spectral state after this pause. In their paper, the authors state that the source spectrum was still hardening until MJD 56730 (our transition state spectrum date is MJD 56703), which is consistent with our determination. It is unclear why the authors considered the source already in the hard state and estimated a higher transition flux and closer distances. 

On the other hand, GRS 1739--278 was discovered initially in its 1996 outburst \citep{paul1996, vargas1997}. The source went into another outburst in 2014 and was observed by \swift~\citep{krimm2014b}. The main 2014 outburst lasted over a year, and the soft-to-hard transition was not observed due to the Sun's obscuration. Following this main outburst, two mini-outbursts were also observed \citep{yan2017}. In the second mini-outburst, observations during the soft-to-hard transition were obtained \citep{yan2017}. As the soft-to-hard transition was observed during that mini outburst, we have used it to estimate a distance of 12.74$^{+3.39}_{-2.42}$\,kpc. In contrast, using interstellar extinction \citet{greiner1996} found the distance to the source in the range of 6-8.5\,kpc. This estimate was later revisited by \citet{yan2017}, where they used an $N_\text{H}$ from \swift/XRT spectral modelling as well as an updated extinction map to estimate a distance of 7.5$\pm2$\,kpc. While their upper limit of 9.5\,kpc is very close to our lower bound of 10.32\,kpc (1$\sigma$), it is still slightly higher. Once again, according to \citet{jonker2004} systematic bias in estimating $N_\text{H}$ through X-ray observations leads to underestimation of distance. This could possibly explain this discrepancy. However, by looking at the distance marginalized distribution in Fig.~\ref{fig:per_sources_distance1} as well as the soft-only estimates in Table~\ref{tab:table2_result} and Fig.~\ref{fig:estimate_vs_lit_soft}, we notice that the higher distance values in our derived combined distance distribution are mainly due to the transition constraint. Due to this fact, we think that another possible explanation for the discrepancy between our estimates is that this source might have its transition luminosity occur at a value lower than the 1-4\% range we used. If we use a lower value of the Eddington fraction, then the transition constraint would shift toward lower distances at the same masses and, in turn, make the combined distribution peak at lower distances. This is entirely possible since we have used a mini-outburst transition flux. The mini-outburst sources observed up till now are a handful, and no systematic studies were made on whether they also follow the empirical correlations found in major outbursts \citep{yan2017}. However, it is worth noting that preliminary analysis shows that the same mechanisms drive them, and thus, we expect them to behave similarly \citep{yan2017}.

\begin{figure*}
\includegraphics{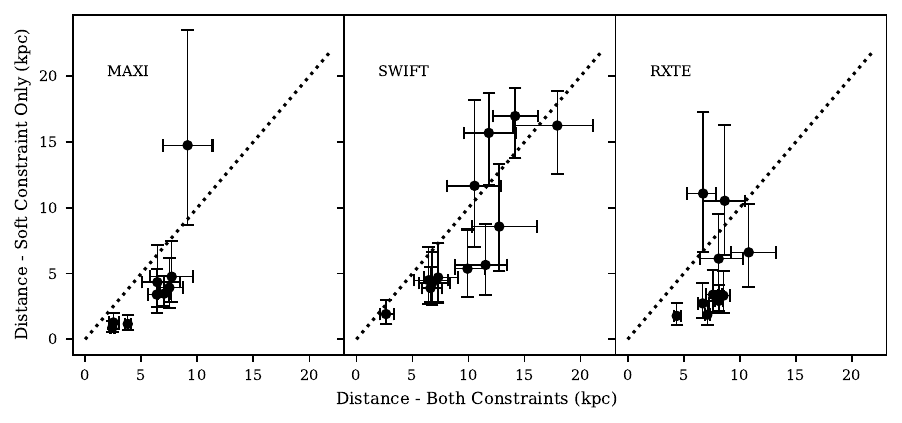}
    \caption{Comparison of the distance constraints obtained using information from both the soft and transition spectra and those obtained using the soft spectra only. A trend is apparent where the soft-only constraints yield lower distances than those constraints using the spectral information from both states.}
    \label{fig:soft_vs_both}
\end{figure*}

While it is clear that we can obtain better constraints from our method if we information from both states, we also want assess the robustness of our distances derived from the soft state spectra alone. In Fig.~\ref{fig:estimate_vs_lit_soft}, the same comparison as above is made, but for the outbursts/sources where we could only find the soft constraint. In the graph, data points from \maxi/GSC and \rxte/PCA suggest that the distance is systematically underestimated by using the soft-only constraint. On the other hand, estimates made using data from \swift/XRT appear to be less affected by the loss of information from the transition state and agree very well with previous estimates from the literature. Nevertheless, when looking at the same source (GX 339-4, black point) that was observed by all instruments, we find consistent evidence that the soft state distance constraints are systematically lower than literature distances. To investigate this further, we also plotted a comparison between our estimates that were found using both constraints and those that were found using the soft-only constraint in Fig.~\ref{fig:soft_vs_both}. The plot shows that apart from six data points at distances larger than 5\,kpc, all other data points indicate that the soft state constraint estimates are systematically lower than estimates found when using both constraints. This is also corroborated by the distance distributions of each source as shown in Fig~\ref{fig:per_sources_distance1} (compare red lines to blue lines). Inspection of the differences between the two estimates (our soft-only estimates versus estimates using both constraints) reveals that distances obtained through the soft-only constraint might be underestimated by up to 70\%. However, exceptions exist for sources located further than 5 kpc, where the soft-state constraint suggests larger distances are needed. This discrepancy could stem from a systematic bias associated with the disk spectrum model we employed. We delve into potential biases with greater detail in Section~\ref{sec:limitations}.

\subsubsection{Consistency Between Instruments}

To investigate whether there are any biases arising from the use of any particular detector, we examined the distance estimates determined from outbursts that observed in multiple instruments. We looked at estimates obtained by using both constraints and through using the soft-only constraint, as the number of overlapping outbursts were small. We found that all estimates are in agreement within their uncertainties. Our findings indicate that there is no detector-specific bias when applying our method to data from \maxi/GSC, \swift/XRT, or \rxte/PCA. This is in line with the expected results since the energy bands used for modelling the data from each instrument are nearly identical. Thus, spectral modelling results are not expected to change significantly.

\subsubsection{Fractional Uncertainty and Dependence on Flux}

So far, we have considered the absolute errors of our distance estimates when determining the quality of the constraints. Using the absolute error, however, ignores that large distances might carry a larger uncertainty as the source flux from these more distant sources would be lower. To be able to compare the quality of measurements better, we first find the average fractional uncertainties of each measurement by calculating the average 1$\sigma$ upper/lower errors, and then we divide this average error by the median distance estimate. Once the fractional uncertainties have been found, we investigate the relationship of the fractional uncertainties with flux. Given that our dataset is small, we calculated the Kendall's tau correlation coefficient between the fractional uncertainty and the log$_{10}$ of each of the soft and the transition fluxes. Both Kendall's tau correlation coefficients were close to zero and had $p$-values of 1.0 and 0.86 for the soft and transition fluxes, respectively. This suggests little to no evidence of an association between the fractional uncertainty and the fluxes. We also tested these correlations for each instrument separately and found similar $p$-values. Having determined that the fractional uncertainties have no detectable dependence on the observed flux. Using all estimates we find a median fractional uncertainty of 0.2, when using spectral information from both the soft and transition states. 

Repeating the above tests for the fractional uncertainties obtained through the soft-only constraint, we calculated the Kendall's tau correlation coefficients separately for each instrument. The Kendall's tau correlation coefficients for \swift~ and \rxte~ indicated there is no association with the soft state log flux. Since the coefficients were close to zero and their $p$-values were equal to 0.84 and 0.28, respectively. On the other hand, the test provided weak evidence of a relationship between \maxi's fractional uncertainties and the soft state flux with a coefficient value of -0.45 ($p$-value = 0.05). So, we conclude that the uncertainties in the distance estimates from \maxi~decrease, albeit weakly, as the source's log$_{10}$ flux increases when only the soft state constraint is found. While for \swift~and \rxte, there is no detectable dependence on the source's flux. We then calculate that the median fractional uncertainty is 0.48 when considering all estimates found using the soft-only constraint. Although more data are needed to provide more robust conclusions regarding the relationship between fractional uncertainty and flux, our analysis here provides a preliminary investigation.

\subsection{Improved Distances}\label{sec:improved-dist}

Our method significantly improved the quality of the distance constraint for several sources. In this section, we take a closer look at sources where we found much more precise distance constraints, for sources which previously only had upper or lower limits on their distances. 

We start by looking at Swift J1910.2-0546, which is a peculiar source because it exhibited two state transitions\footnote{We used the first soft-to-hard transition in our estimation} \citep{nakahira2014, reis2013}. Applying our method to data from the 2012 outburst of Swift J1910.2-0546 obtained by \maxi/GSC, we were able to find a precise distance estimate to the source of 3.8$^{+0.23}_{-0.25}$\,kpc. The only available estimate for this source in the literature was made by \citet{nakahira2014} using the simple approach of our method, which assumes a non-rotating black hole and enforces discrete estimates on inclination. The authors only reported a lower bound of 1.7\,kpc. Since the authors used the same approach as our method, the consistency between our measurements is not surprising. Another source is XTE J1652-453; the distance had been previously estimated to be $\geq5$\,kpc \citep{reynolds2009}, using the spectrum of the companion star. Applying our method to data from \rxte/PCA provides the most precise constraint on the distance of XTE J1652-453, as we calculated a distance of 6.69$^{+0.54}_{-0.26}$\,kpc. Moreover, using our method on data from \rxte/PCA of the 1998 outburst of XTE J1748-288 and the 2008 outbursts of Swift J1842.5-1124, we confined their distances to 8.66$^{+1.98}_{-1.84}$ kpc and 8.12$^{+1.67}_{-2.2}$\,kpc, respectively. The prior estimate of XTE J1748-288 was $\geq$8\,kpc \citep{hjellming1998}, which was obtained using an H\,{\sc i} absorption measurement. While the best estimate of Swift J1842.5-1124 of $>5$\,kpc \citep{zhang2022} was found by utilizing radio and X-ray correlations.

Additionally, using the \maxi/GSC data of Swift J1727.8-1613 in its ongoing 2023 outburst, we have utilized the soft state spectra to estimate a distance of 1.52$^{+0.85}_{-0.61}$\,kpc. Although the source is currently still in outburst, the HR data suggest that the source is in the soft state \citep{bollemeijer2023}. This estimate perfectly agrees with the current only estimate in the literature of 1.5\,kpc using flux scaling arguments \citep{veledina2023}. We point that our estimate is obtained through the soft state information only, so the distance might be underestimated by up to $\sim 70\%$ (see Section~\ref{sec:comp_to_prev}). 

In general, we have improved the distance estimates for 10 sources, achieving up to a 45\% decrease in fractional uncertainty for Swift J1910.2-0546 compared to previous estimates (see Table~\ref{tab:comp_table}).

However, it is important to note that, while our pragmatic framework has found much tighter constraints than many previously estimated distances, we must emphasize that our errors are statistical and may be subject to systematic biases. In our discussion in Section~\ref{sec:limitations}, we will attempt to identify these biases and suggest ways to refine our framework to accommodate them in future work or in form of adjustments to the pipeline.

\subsection{Overall Distributions}

\subsubsection{Distance and Spatial Distributions}

In Fig.~\ref{fig:overall_dist_1d}, we plotted the overall distance distribution obtained by adding the probability distributions from each source in our sample, visualizing the probability of finding BH-LMXBs as a function of distance. It is tempting to correlate the multi-peak structure of the observed distribution with the Milky Way's spiral structure. However, since these distances are measured directly from the sun to the respective sources, they do not necessarily provide insights into the Galaxy's overall structure. Thus, we cannot infer any relationship between the Milky Way's structure and this distance distribution. Furthermore, this distribution was obtained considering only the statistical uncertainties. Including systematic errors (see Section~\ref{sec:limitations}) would possibly smear the spiky shape reminiscent of the spiral arms over-densities.

On the other hand, the spatial distribution of sources in our sample, as shown in Fig.~\ref{fig:spatial_distribution}, shows that the sources are likely to be located near spiral arms or the Bulge. Comparing this observed distribution to a similar observed distribution as in \citet{corral-santana2016,gandhi2019}, we found that our inferred distribution is consistent with their distribution. In particular, all distributions find that BH-LMXBs are concentrated towards the direction of the Galactic Bulge, with distances indicating positions at or very near the Galactic center. We also find a small number of sources located further away from the Galactic center. Careful investigations on the selection effects affecting our observed distributions will be the subject of one of our future paper.

\subsubsection{Mass}
On the other hand, looking at the posterior mass distribution in Fig.~\ref{fig:overall_mass_1d}. We find that the probability density peak appears to have moved slightly higher (to 8.7 \ms~from 7.8 \ms) when compared to our prior guess. Thus, this indicates that we have gained new information on the most frequent or likely mass values, which are now slightly higher than before. Additionally, there's a notable increase in probability density within the 10 to 12.5 \ms~range. This observed mass distribution appears to reaffirm the observed lower mass gap between the 3 and 5 \ms \citep{ozel2010, farr2011, kriedberg2012}. This is expected since we use the distribution found by \citet{ozel2010}, which is our prior mass guess. But since there was no \textit{a priori} constraint on the direction of the information gain in the posterior mass distribution (i.e. we could have found sources where both constraints only agree at much lower masses), we argue that our results contribute to the evidence of the observed lower mass gap. Nonetheless, we cannot exclude that it is due to an observational bias. Further detection of new BH-LMXBs and determining their masses would pave the way to understanding these possible biases.  

\subsection{Limitations and Biases}\label{sec:limitations}
In the previous sections, we have demonstrated that our framework has the potential to provide robust estimates for most BH-LMXBs. However, we must discuss a number of limitations and possible biases that may arise from the assumptions we made. 

First, in our disk model, we have assumed a color correction factor $f_{\text{col}}=1.7$. This correction factor is needed because the emission from the disk is not a perfect blackbody radiation \citep{ebisuzaki1984}. \citet{shimura1995} argued that the spectra produced could be approximated by a blackbody with maximum temperature shifted to higher temperatures by $f_{\text{col}}$. By numerically solving the equations of the vertical structure of the disk, \citet{shimura1995} found this factor should be in the range 1.5-1.9, with 1.7 being the most acceptable approximation. Since $f_{\text{col}}$ enters in the distance/mass relation found by modelling the soft state (equation~\ref{eq:norm_ezdiskbb_corrections}), it affects our distance estimations. Several studies \citep{merloni2000, davis2005} have shown that there is still large uncertainty in the ranges that $f_{\text{col}}$ can take. Nonetheless, most theoretical models suggest it should be in the narrow range of 1.4-2.0 \citep{davis2005,davis2019}. Our assumption of $f_{\text{col}}=1.7$ was motivated by empirical evidence, which found that most systems are consistent with the value of 1.7 \citep{reynolds2013}. We also note that for the disk-dominated state (soft spectral state) with disk temperatures of $<1$\,keV (typical of all our spectral fits), the latest theoretical models predicated that $f_{\text{col}}$ should be 1.7-1.8 \citep{davis2005, reynolds2013, davis2019}. Furthermore, the higher values $f_{\text{col}}$ are only predicted for systems, which are accreting at high Eddington fractions ($> 30$\%) \citep{davis2019}. Observational evidence hints that these systems are rare \citep{davis2019}. Additionally, relativistic disk models\footnote{The relativistic corrections were found by utilizing the relativistic disk model \texttt{kerrbb} (see Section~\ref{sec:method-correc}).} already utilize the Novikov-Thorne disk model \citep{nt1973}, which assumes thin disks with Eddington ratios less than 30\%. Therefore, considering higher values seems unwarranted as it is not consistent with the models we use. Notwithstanding, we identify this as a possible bias source. In preliminary analysis, where we allowed $f_{\text{col}}$ to vary in the range of 1.4 to 2.0 for MAXI J1820+070 using \swift/XRT data, we have found minimal change in the distance from 2.64$^{+0.71}_{-0.56}$\,kpc to 2.63$^{+0.71}_{-0.55}$\,kpc. This very minimal change is almost certainly associated with the tight soft-to-hard transition constraint of this source (see Fig.~\ref{fig:per_sources_mass_distance1}) which means that the overall estimate is not highly sensitive to changes in $f_{\text{col}}$ . Thus, being able to draw conclusions on this matter would require analysis of multiple sources with varying degrees of narrowness for the soft-to-hard transition constraint. Since including the uncertainty of $f_{\text{col}}$ in addition to other certainties in our current implementation of the method renders the probability calculation very computationally intensive, we have reserved full investigation of the effect of $f_{\text{col}}$ on our estimations for future work. 

Another source of bias is the use of the \citet{ozel2010} mass prior. We argue that our motivation was purely observational since several studies have reinforced this prior \citep{farr2011, kriedberg2012}. But we also note that there are some possible physical mechanisms that can explain the observed distribution, especially for the low mass gap at 2-5\ms, such as natal kicks that favour the formation of X-ray binaries in a certain mass range \citep{fryer2001, fryer2012}. We also note that using both the soft and transition constraints decreases the effect of this bias. Because agreement between both constraints is required for updating the prior, as shown in Fig.~\ref{fig:overall_mass_1d}. 

The use of the 1-4\%$L_{\text{Edd}}$ assumption of the soft-to-hard transition luminosity \citep{maccarone2003} is also another source of bias. It has been suggested that the state transition luminosity might depend only on the dimensionless viscosity parameter, and the observational evidence suggests that this parameter does not appear to change significantly between different sources \citep{maccarone2003, vahdat2019}. As we mentioned in the methods section, several studies found observational evidence for the same range proposed by \citet{maccarone2003}. Most recently, a study by \citet{wang2023} investigated the transition luminosities again using data from \swift/XRT and \nicer/XTI. The authors found that the power law flux is tightly constrained around 1.3\%. Moreover, the authors point to the importance of the methodology for determining the state transition. They argue that the use of disk recession (when disk normalization begin to increase) would be simpler to verify when analyzing spectra that have been observed in the soft bands (0.5-10\,keV). Given that we used the common approach of determining the state transition through HR analysis, this is yet another source of bias because the definitions of HR can vary between studies. Nevertheless, we believe that the 1-4\% range we considered is large enough to include the uncertainties in determining the state transition dates. 

A limitation of our method is the presence of insufficient spectral data at the state transition. We note that we could not obtain a distance estimate for most of the sources from our initial sample because they suffered the absence of viable observations at the state transition. Additionally, our most robust measurements are obtained using both constraints. Fortunately, in the future, continuous monitoring from missions such as \maxi/GSC and \swift/BAT would reduce cases of fewer observations in the transition state.

Finally, a systematic bias in our estimates could be from the assumption of prograde spins only ($a>0$). Strong observational evidence shows that the majority of discovered BH-LMXBs have prograde spins \citep{reynolds2021}. However, we have to note that there are some exceptions where it has been suggested that a retrograde spin is most likely \citep[e.g.][]{reis2013, sandeep2020}. This would affect the soft state constraint and would increase the probability density in the direction of higher distances\footnote{Since $r_{\text{in}}$ would increase, so for the same disk normalization, $M$, $i$, and spin, the distance would increase}. To further investigate what impact the inclusion of retrograde spin would have on the calculated distances, we recomputed the distances for six sources, employing a flat prior for \(a\) ranging from -0.999 to -0.998. The sources analyzed include GRS 1739-278, with a recalculated distance of \(14.05^{+3.2}_{-3.4}\) kpc; MAXI J1348-630, at \(2.52^{+0.31}_{-0.45}\) kpc; Swift J1910.2-0546, at \(3.76^{+0.37}_{-0.51}\) kpc; Swift J1842.5-1124, at \(8.7^{+1.99}_{-2.06}\) kpc; XTE J1650-500, at \(4.66^{+0.62}_{-0.93}\) kpc; and MAXI J1820+070, at \(3.48^{+0.79}_{-0.77}\) kpc based on \maxi/GSC data and \(2.8^{+0.64}_{-0.65}\) kpc based on \swift/XRT data. Comparison with the results listed in Table~\ref{tab:table1_result} reveals that, except for Swift J1910.2-0546, the median distances for all sources have increased. Additionally, the fractional uncertainty in most of the distances has increased. This is a consequence of allowing for retrograde spin, which widens the soft state constraint and shifts its peak to larger distances. This alteration further results in a greater overlap between the soft and transition distributions at larger distances, leading to an increase in the 1\(\sigma\) statistical errors of the combined distribution.

As a final note, we point to the fact that our aim was to use the best current theoretical and observational knowledge to obtain reliable estimates of the distances to BH-LMXBs. Thus, we have made a number of pragmatic assumptions to be able to achieve this goal. We also emphasise that our versatile framework could be easily modified to address some of these possible biases if more observational evidence surfaces against the assumptions we used in our work. So, the framework's robustness lies not only in the current implementation but also in its ability to evolve with the latest observational evidence and theoretical understanding. For instance, further modification to our method would include the possibility of accounting for $f_{\text{col}}$ uncertainties, retrograde spin, or a spin prior based on the observed BH-LMXB population.

\section{Conclusion}\label{sec:conc}
We have analyzed the light curves of 50 transient low-mass X-ray binaries selected from the catalogues of \citet{tetarenko2016}, \citet{corral-santana2016}, and \citet{alabarta2021} using archival data from \maxi/GSC, \swift/XRT and/or \rxte/PCA. We further selected sources that had an identifiable soft state and excluded sources with very poor spectral fits ($\chi_{\text{red}}>1.4$). Consequently, this led to a final sample of 26 sources. We performed X-ray spectral analysis of 39 soft-state spectra and 27 soft-to-hard transition spectra for a total number of 34 unique BH-LMXB outbursts. The main results from our study can be summarized as follows:

\begin{enumerate}
    \item We have developed a robust framework to constrain the black hole BH-LMXB distances from their X-ray spectra, considering general relativistic effects and other physical parameter uncertainties.
    \item Using this framework, we constrained the distances to 26 sources, often improving their existing constraints.
    \item Comparison to previous distance estimates shows good agreement. In particular, our estimates are in good agreement with the highly accurate estimates obtained through Gaia parallax measurements and H\,{\sc i} absorption methods.
    \item We have not detected any instrument-specific biases in the distance estimates obtained using data from \maxi/GSC, \swift/XRT, or \rxte/PCA.
    \item We have found that estimates obtained using both the soft and transition spectral information have a median uncertainty of 20\% while estimates obtained using only the soft state spectrum have a median uncertainty of about 50\%.
    \item We have created and published a \texttt{python} pipeline that can be used to calculate the distance/mass probability densities and has implemented an online tool based on the results of this study that can be readily used to quickly calculate the distance estimates.
    \item Using our distance estimates, we found an overall probability distribution of distances (Fig.~\ref{fig:overall_dist_1d}) from the sun with peaks at 2.4, 4, 7.05, and 8.1 kpc. Investigating the spatial distribution of the sources in the Milky Way plane projection (Fig.~\ref{fig:spatial_distribution}), we found that most of the sources are directed toward the Bulge and near the spiral arms.
    \item The posterior mass distribution found by requiring the estimates from the soft state and the transition state must agree on the true value of distance, contributes to the evidence for the 3-5\ms~mass gap \citep[e.g.][]{ozel2010,fryer2012}. 
    \end{enumerate}
We note that our distance estimation framework can be applied on data from other X-ray detectors such as \nicer. Nevertheless, due to the low amount of outbursts observed with \nicer~compared to other the instruments we used, we have reserved analysis using \nicer~ for future projects when more data is available.

\section*{Acknowledgements}

We deeply appreciate the insightful comments and constructive suggestions provided by the anonymous referee, which have significantly enhanced the quality and clarity of our paper. Y.A. gratefully acknowledges NASA for support under the \swift~Guest Investigator Grant 80NSSC23K1148. Y.A. also thanks Greg Salvesen, Jonah M Miller and the Los Alamos National Laboratory (LANL) for making the general relativistic correction factors data public (\url{https://github.com/gregsalvesen/bhspinf}) which was valuable in the production of this work. This work made use of Astropy:\footnote{\url{http://www.astropy.org}} a community-developed core Python package and an ecosystem of tools and resources for astronomy \citep{astropy:2013, astropy:2018, astropy:2022}. The authors also acknowledge the developers of \textit{NumPy} \citep{harris2020array}, \textit{PyTorch} \citep{pytorch}, \textit{SciPy} \citep{scipy}, and \textit{boost-histogram} \citep{bh}.

 \section*{Data Availability}
This research made use of \swift/XRT data, which are publicly available and can be accessed through the NASA's HEASARC (High Energy Astrophysics Science Archive Research Center) database: \url{https://heasarc.gsfc.nasa.gov/cgi-bin/W3Browse/w3browse.pl} and \url{https://www.swift.ac.uk/user_objects/}. Observational data from the Monitor of All-sky X-ray Image/Gas Slit Camera (MAXI/GSC) were also employed in this study. The \maxi/GSC data are freely accessible via the MAXI team's website hosted by the \textit{RIKEN}, \textit{JAXA}, and the \maxi~team \url{http://maxi.riken.jp/mxondem/} and part of the HEAsoft package \url{https://heasarc.gsfc.nasa.gov/docs/software/heasoft/}. Additionally, data from the Rossi X-ray Timing Explorer/Proportional Counter Array (\rxte/PCA) were used. These data are available in the public domain and can be retrieved from the NASA's HEASARC database:  \url{https://heasarc.gsfc.nasa.gov/cgi-bin/W3Browse/w3browse.pl} for the spectra and \url{https://www.isdc.unige.ch/integral/heavens} for light curves. Furthermore, the pipeline used to generate the distance/mass probability densities is available on GitHub at \url{https://github.com/ysabdulghani/lmxbd}, and the scripts for data reduction and preprocessing can be provided upon request.
 


\bibliographystyle{mnras}
\bibliography{references}



\appendix
\section{Spectral Modeling Table}
\begin{landscape}
\begin{table}
\caption{This table contains the information of each spectrum modelled in this work. We show the source name, outburst year, and instrument. We also show the selected spectra dates for the soft state spectrum and, if available, the soft-to-hard transition spectrum (See Section~\ref{sec:state_identify} for more details on selection method). We list the $N_\text{H}$ and its reference, if fixed. The $\chi^2_{\text{red}}$ for the fitted soft state spectrum and, if applicable, the soft-to-hard transition spectrum are also listed. Finally we specify the model used in the fitting (see Section \ref{sec:spec_model} for the definition of models). The sources are ordered in alphabetical order from top to bottom.}
\label{tab:fit_table}
\begin{filecontents*}{tables/fits.csv}
Name,Outburst Year,Instrument,softDates,softChiSquared,softDOF,softRedChiSquared,softModel,transDates,transChiSquared,transDOF,transRedChiSquared,transModel,nH,nH-ref-num,nH-ref
GRS 1739-278,2015,XRT,57250,489.247,508,0.96,Model 1,57266,510.0076,470,1.09,Model 1,2,1,
GX 339-4,2010/2011,MAXI,55520-55521,87.837,68,1.29,Model 1,55597-55606,47.2478,55,0.86,Model 1,0.5,2-3,
,2007,XRT,54216,607.684,502,1.21,Model 1,54240,607.2898,596,1.02,Model 1,Fitted,,
,2020,MAXI,58873-58877,81.119,68,1.19,Model 1,58930-58936,42.05494,50,0.84,Model 1,Fitted,,
,2020,XRT,58915,452.697,398,1.14,Model 1,58953,502.7943,507,0.99,Model 1,Fitted,,
,2015,XRT,57111,570.138,516,1.10,Model 1,57263,592.7372,605,0.98,Model 1,0.5,2-3,
,2015,MAXI,57176-57177,51.869,51,1.02,Model 1,57253-57264,46.62652,57,0.82,Model 1,0.5,2-3,
,2004/2005,RXTE/PCA,53309,24.921,33,0.76,Model 2,53466,34.53837,34,1.02,Model 2,0.5,2-3,
,2010/2011,RXTE/PCA,55430,23.033,34,0.68,Model 2,55594,32.93435,34,0.97,Model 2,0.5,2-3,
H 1743-322,2010,MAXI,55433-55437,75.095,70,1.07,Model 1,55449-55455,53.87373,59,0.91,Model 1,2.2,4-5,
,2013,XRT,56524,649.503,634,1.02,Model 1,,,,,,Fitted,,
IGR J17091-3624,2011/2012,XRT,55703,646.487,642,1.01,Model 1,56041,264.0393,301,0.88,Model 1,Fitted,,
,2016,XRT,57508,564.360,564.3595723,0.94,Model 1,57594,280.7584,284,0.99,Model 1,Fitted,,
MAXI J0637-430,2019/2020,XRT,58823,546.864,392,1.40,Model 1,58858,367.235,338,1.09,Model 1,0.04,6,
MAXI J1305-704,2012,XRT,56087,488.201,451,1.08,Model 1,56098,452.479,460,0.98,Model 1,Fitted,,
MAXI J1348-630,2019,MAXI,58561,110.199,102,1.08,Model 1,58600-58602,97.63783,113,0.86,Model 1,0.86,7,
MAXI J1543-564,2011,XRT,55800,273.295,285,0.96,Model 1,,,,,,Fitted,,
,2011,RXTE/PCA,55759,23.387,34,0.69,Model 2,,,,,,0.9,8,
MAXI J1659-152,2010,XRT,55488,700.237,604,1.16,Model 1,,,,,,Fitted,,
,2010,RXTE/PCA,55477,37.014,36,1.03,Model 1,55502,35.57144,36,0.99,Model 1,0.18,9,
MAXI J1727-203,2018,MAXI,58277-58285,34.600,39,0.89,Model 1,,,,,,0.44,10,
MAXI J1803-298,2021,MAXI,59369-59375,60.264,57,1.06,Model 1,,,,,,0.3,11-12,
MAXI J1820+070,2018,XRT,58317,575.826,494,1.17,Model 1,58387,565.661,507,1.12,Model 1,Fitted,,
,2018,MAXI,58371-58372,58.657,59,0.99,Model 1,58382,56.986,60,0.95,Model 1,0.15,13,
MAXI J1828-249,2013,MAXI,56581-56583,68.279,65,1.05,Model 1,56620-56626,47.32135,47,1.01,Model 1,0.2,14,
Swift J1539.2-6227,2009,RXTE/PCA,54846,18.602,34,0.55,Model 2,54867,38.01898,34,1.12,Model 2,0.35,15,
Swift J1727.8-1613,2023,MAXI,60281,62.401,63,0.99,Model 1,,,,,,0.4,16,
Swift J1842.5-1124,2008,RXTE/PCA,54741,43.452,34,1.28,Model 2,54763,34.48866,34,1.01,Model 2,0.3,17,
Swift J1910.2-0546,2012,MAXI,56084-56085,68.281,54,1.26,Model 1,56132-56135,83.07185,81,1.03,Model 1,0.35,18,
XTE J1650-500,2001,RXTE/PCA,52200,25.557,33,0.77,Model 2,55235,38.7507,32,1.03,Model 2,0.67,19,
XTE J1652-453,2009,RXTE/PCA,55042,23.750,34,0.70,Model 2,55083,25.33797,34,0.75,Model 2,6.7,20,
XTE J1720-318,2003,RXTE/PCA,52669,16.838,34,0.50,Model 2,,,,,,1.3,21,
XTE J1748-288,1998,RXTE/PCA,50982,26.647,38,0.70,Model 2,51007,33.29914,38,0.88,Model 2,Fitted (High),,
XTE J1752-223,2010,RXTE/PCA,55255,37.225,34,1.09,Model 2,55284,41.28423,33,1.25,Model 2,0.5,22,
,2010,MAXI,55218,50.636,59,0.86,Model 1,,,,,,0.5,22,
XTE J1817-330,2006,RXTE/PCA,53838,26.666,33,0.81,Model 2,,,,,,0.12,23,
XTE J1818-245,2005,RXTE/PCA,53639,22.065,33,0.67,Model 2,,,,,,0.54,24,
XTE J1859+226,1999/2000,RXTE/PCA,51513,30.552,33,0.93,Model 2,,,,,,0.6,25-26,
XTE J1908+094,2013/2014,XRT,56608,603.690,548,1.10,Model 1,56705,234.1606,284,0.82,Model 1,Fitted,,
\end{filecontents*}
\pgfplotstabletypeset[col sep=comma,
columns/Name/.style={string type,column name = Source,column type=l},columns/Outburst Year/.style={string type},columns/Instrument/.style={string type,column name=Instrument},columns/nH/.style={string type,column name={\makecell[t]{
   $N_{\text{H}}$ \\ ($\times 10^{22}$ cm$^{-2}$)} }},columns/nH-ref-num/.style={string type,column name={\makecell[t]{
   $N_{\text{H}}$ \\ Ref.}},string replace={}{--}},columns/softRedChiSquared/.style={string type,column name= $\chi^2_{\text{red,soft}}$,string replace={}{--}},columns/transRedChiSquared/.style={string type,column name = $\chi^2_{\text{red,trans}}$,string replace={}{--}},columns/softModel/.style={string type,column name = Soft Spectrum Model,string replace={}{--},string replace={Model 1}{S1},string replace={Model 2}{S2}},columns/transModel/.style={string type,column name = {\makecell[t]{Transition Spectrum \\ Model}},string replace={}{--},string replace={Model 1}{T1},string replace={Model 2}{T2}}
,columns/softDates/.style={string type,column name =  {\makecell[t]{
   Soft Spectrum \\ Date(s) (MJD)}}},columns/transDates/.style={string type,column name= {\makecell[t]{
   Transition Spectrum \\ Date(s) (MJD)}},string replace={}{--}},every head row/.style={before row=\toprule,after row=\midrule}, every row no 0/.style={before row=\hline}, every row/.style={before row=\hline},
every last row/.style={after row=\bottomrule}
,columns={Name,Outburst Year,Instrument,nH,nH-ref-num,softDates,softModel,softRedChiSquared,transDates,transModel,transRedChiSquared}
]{tables/fits.csv}
\begin{tablenotes}[flushleft]
\small
\item \textbf{References.} (1) \citet{wang2018}, (2) \citet{mendez1997}, (3) \citet{yang2023} (4) \citet{marco2016}, (5) \citet{chand2021}, (6) \citet{jia2023}, (7) \citet{tominaga2020}, (8) \citet{kennea2011}, (9) \citet{kalamkar2011}, (10) \citet{alabarta2020},  (11) \citet{bult2021},  (12) \citet{homan2021} (13) \citet{shidatsu2019}, (14) \citet{oda2019}, (15) \citet{krimm2011}, (16) \citet{draghis2023b}, (17) \citet{zhao2016}, (18) \citet{degenaar2014}, (19) \citet{tomsick2004}, (20) \citet{hiemstra2011}, (21) \citet{cadolle2004}, (22) \citet{ratti2012}, (23) \citet{rykoff2007}, (24) \citet{cadolle2009}, (25) \citet{markwardt1999}, (26) \citet{dalfiume1999}
\end{tablenotes}
\end{table}
\end{landscape}

\section{Probability Distributions}
\begin{figure*}
\includegraphics{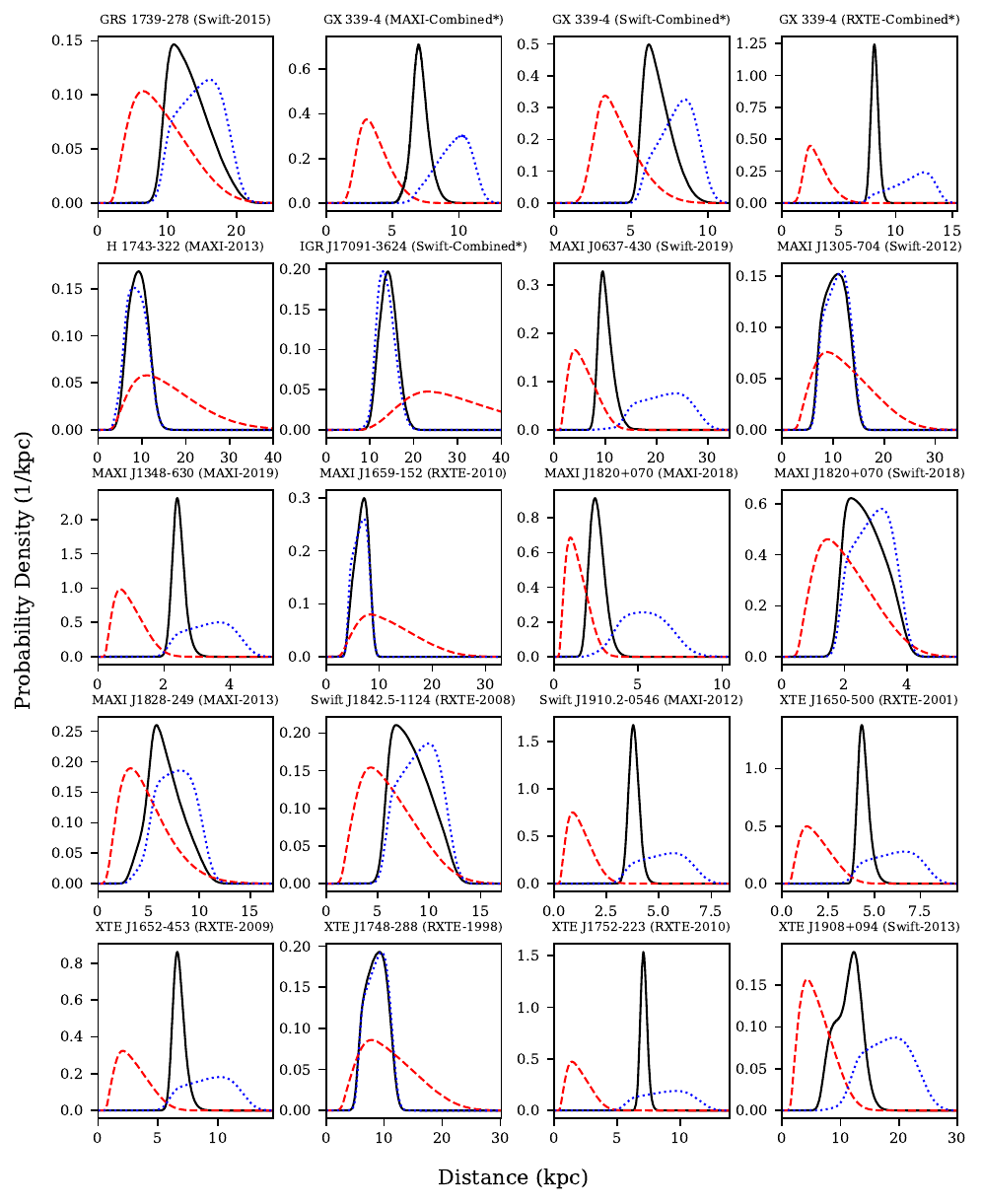}
\caption{This figure shows the marginalized distance probability distribution for all of the outbursts in our sample that both constraints. The red line shows the probability distribution from only the soft state constraint, while the blue line shows the probability distribution from only the transition luminosity. The black line shows the probability distribution that results by combining both constraints such that both are assumed to be true, equation ((\ref{eqn:combined_soft_trans_pdf})). For sources with multiple outbursts, we only use the combined probability distribution. * See Table~\ref{tab:table1_result} for outbursts dates used to obtain the combined estimate.}
\label{fig:per_sources_distance1}
\end{figure*}

\begin{figure*}
\includegraphics{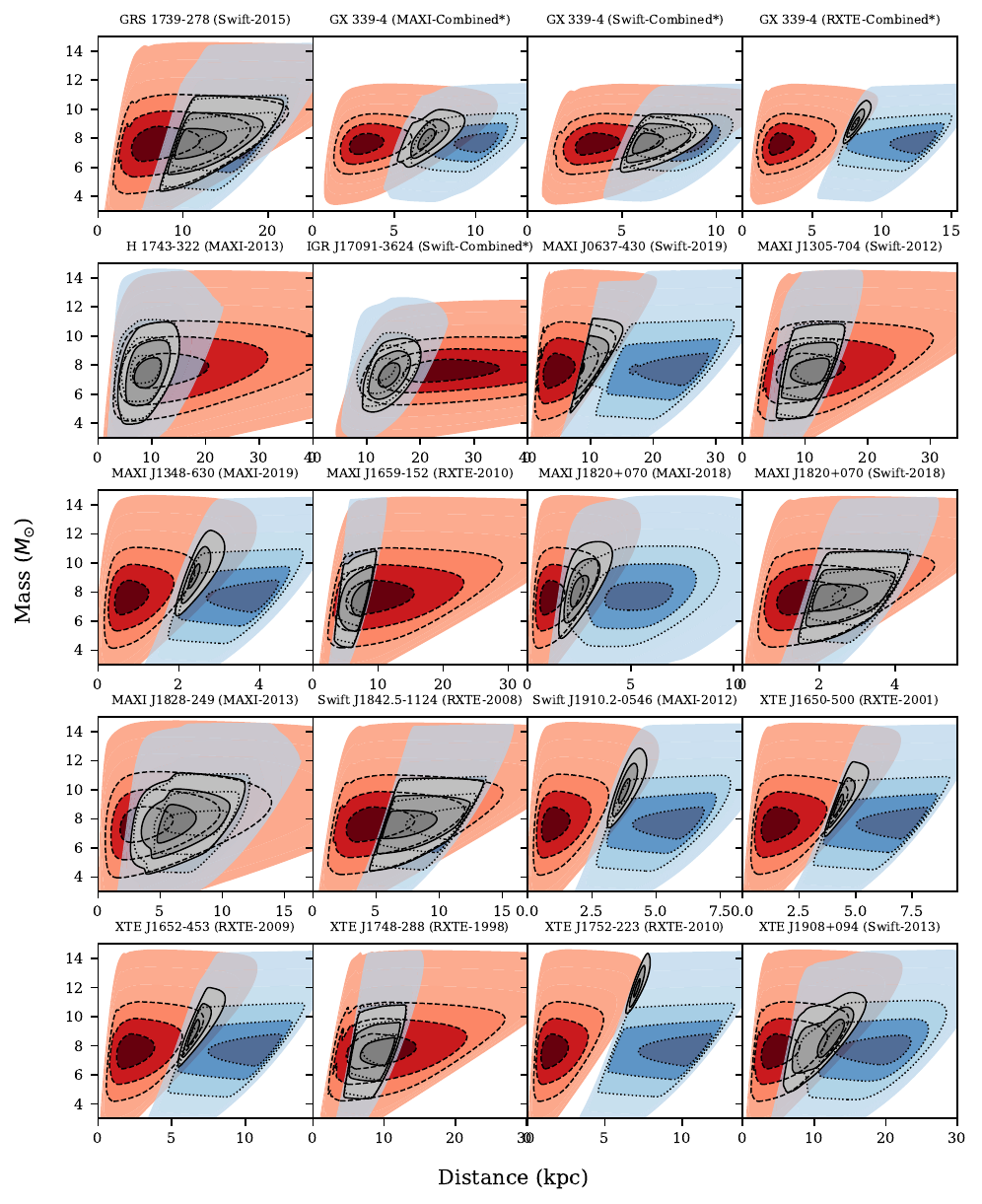}
\caption{This figure shows the joint mass-distance probability distribution for all of the outbursts in our sample using both constraints. The red region shows the probability distribution from only the soft state constraint while the blue region show the probability distribution from only the transition luminosity. The gray region shows the combined probability distribution that results when both distributions are assumed to be true, equation (\ref{eqn:combined_soft_trans_pdf}). All contours lines are 1$\sigma$, 2$\sigma$, and 3$\sigma$ regions. For sources with multiple outbursts we only use the combined estimated probability distribution. * See Table~\ref{tab:table1_result} for outbursts dates used to obtain the combined estimate.}
\label{fig:per_sources_mass_distance1}
\end{figure*}

\begin{figure*}
\includegraphics{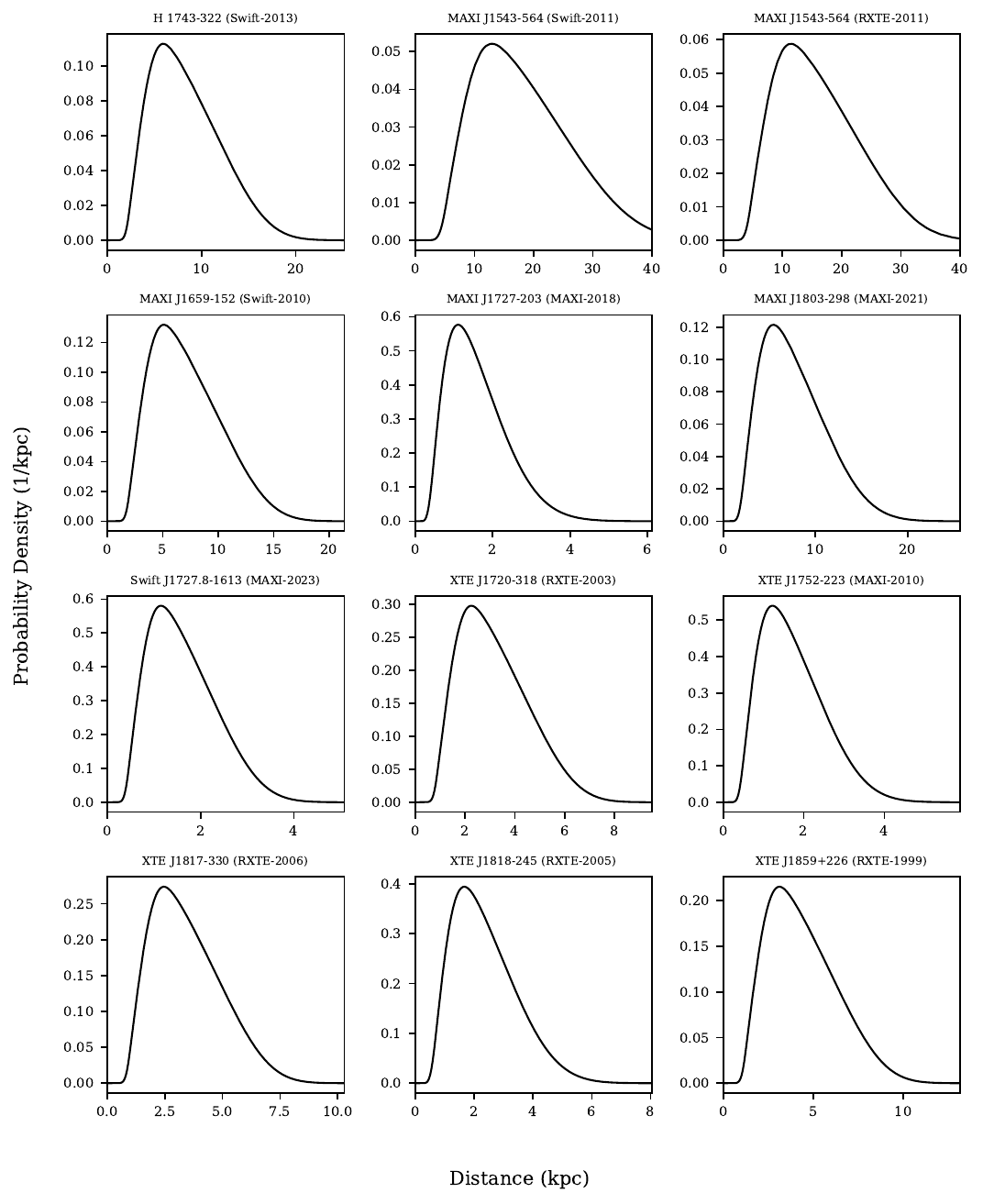}
\caption{This figure shows the distance probability distribution for sources/outbursts from our sample with only spectral information from the soft state.}
\label{fig:per_sources_distance_soft}
\end{figure*}

\section{Comparison to Previous Estimates Table}
\begin{table*}
\caption{In this table we show the best estimated distance (this work) for each source in our sample as well as the most accurate distance from the literature. We also show the distance method used to obtain the literature estimate. Sources are ordered alphabetically from top to bottom.}
\label{tab:comp_table}
\begin{filecontents*}{tables/compTable.csv}
Name,Outburst(s) Year,Outburst(s) Start-End Time (MJD),insturment,literature-mass,literaturemass-lowerr,literaturmass-uperr,literaturemass-ref,literaturespin,literaturespin-lowerr,literaturespin-uperr,literaturespin-ref,previous estimate method,lit-dist-method-short,prevmethod-dist,prevmethod-lowerr,prevmethod-uperr,prevmethod-ref,best-estimate-med,best-estimate-low,best-estimate-upper,inc-low,inc-high,inc-ref,ra,dec,nH,nH ref
GRS 1739-278,2014,56700-56996,maxi and xrt,none,none,none,,none,none,none,,Interstellar Extinction,D,7.5,2,2,\citep{yan2017},12.74,2.42,3.39,33,60,wang2018,265.6668,-27.748,2, wang2018
GX 339-4,2004-2005,53054-53515,rxte,7.5,3.5,3.5,\citep{zdziarski2019},0.95,0.08,0.02,\citep{parker2016},Reflection \& Continuum-Fitting,C,8.4,0.9,0.9,\citep{parker2016},8.13,0.3,0.34,46,70,{zdziarski2004,heida2017},255.7058,-48.7897,0.5,{mendez1997,yang2023}
H 1743-322,2013,56506-56567,maxi and xrt,10,5,5,\citep{steiner2012},0.25,0.5,0.5,\citep{steiner2012},Proper Motion (Jet),I,8.5,0.8,0.8,\citep{steiner2012},9.14,2.22,2.22,70,89,chakrabarti2019,266.565,-32.2336,2.2,{marco2016,chand2020}
IGR J17091-3624,2011-2013,55590-56412,xrt,12.75,0.95,0.95,\citep{Iyer2015},none,none,none,none,Hard-to-soft Transition Luminosity,H,14,3,3,\citep{rodriguez2011},14.16,1.94,2.03,30,40,xu2017,257.2817,-36.4071,Fitted (1.4),
MAXI J0637-430,2019,58790-58950,xrt,8.5,3.5,3.5,\citep{jana2021},none,none,none,none,Hard-to-soft Transition Luminosity,H,8.7,2.3,2.3,\citep{soria2022},9.91,1.1,1.57,58,70,lazar2021,99.0983,-42.8678,0.0439,jia2023
MAXI J1305-704,2012,56009-56190,maxi and xrt,8.9,1,1.6,\citep{matasanchez2021},none,none,none,,Companion Star,E,7.5,1.4,1.8,\citep{matasanchez2021},10.54,2.42,2.35,60,80,sanchez2021,196.7304,-70.4514,0.18,feng2023
MAXI J1348-630,2019,58509-58684,maxi,9.1,1.2,1.6,\citep{jana2020_j1348},0.78,0.04,0.04,,HI Absorption,B,2.2,0.6,0.5,\citep{chauhan2021a},2.42,0.16,0.2,0,60,tominaga2020,207.0531,-63.2749,0.86,tominaga2020
MAXI J1543-564,2011,55653-55832,xrt,13,0.4,1,\citep{chatterjee2016},none,none,none,none,X-ray Decay Time Scale,F,8.5,0,100,\citep{stiele2012},15.12,6.01,8.39,0,89,none,235.8223,-56.4136,Fitted (2.2),kennea2011
MAXI J1659-152,2010-2011,55456-55614,maxi and rxte,5.4,2.1,2.1,\citep{torres2021},-0.28,0.72,0.72,\citep{feng2022},Interstellar Extinction,D,8.6,3.7,3.7,\citep{kuulkers2013},6.73,1.46,1.19,60,80,kuulkers2013,254.757,-15.258,0.18,kalamkar2011
MAXI J1727-203,2018,58274-58400,maxi,none,none,none,,none,none,none,,Soft state \& Soft-to-hard Transition,J,5.9,0,100,\citep{wang2022},1.47,0.61,0.9,0,89,alabarta2020,261.971,-20.389,0.44,alabarta2020
MAXI J1803-298,2021,59336-59450,maxi,none,none,none,,,,,,Soft state \& peak lumonsity ,O,6,0,100,\citep{shidatsu2022},7.19,2.9,4.13,,,,270.923,-29.804,0.3,{bult2021,homan2021}
MAXI J1820+070,2018,58190-58410,maxi and xrt,8.48,0.72,0.79,\citep{torres2020},none,none,none,,Gaia Parallax,A,2.66,0.52,0.85,\citep{gaiaed3,tetarenko2021},2.52,0.4,0.49,60,70,shidatsu2019,275.0914,7.1854,0.15,shidatsu2019
MAXI J1828-249,2013,56580-56640,maxi,none,none,none,,none,none,none,,Soft state \& Soft-to-hard Transition,J,7,0,0,\citep{oda2019},6.44,1.41,1.99,0,89,oda2019,277.242,-25.0294,0.2,oda2019
Swift J1539.2-6227,2008-2009,54792-54966,rxte,none,none,none,none,none,none,none,none,Radiatied energy and rise/decay time scale correlations,N,9.32,1.53,1.53,\citep{yan2015},10.8,1.62,2.4,70,89,krimm2011,234.7998,-62.4673,0.35,krimm2011
Swift J1727.8-1613,2023,,maxi and xrt,none,none,none,none,none,none,none,none,Flux Scaling,M,1.5,0,0,\citep{veledina2023},1.52,0.61,0.85,30,60,\citep{veledina2023},261.9417,-16.21472,Fitted,
Swift J1842.5-1124,2008-2009,54630-54857,rxte,none,none,none,none,none,none,none,none,Radio \& X-ray Correlation,K,5,0,100,\citep{zhang2022},8.12,1.67,2.2,0,89,none,280.5727,-11.4177,0.3,zhao2016
Swift J1910.2-0546,2012-2013,56060-56329,maxi,none,none,none,,none,none,none,,Soft state \& Soft-to-hard Transition,J,3.5,1.8,1.8,\citep{nakahira2014},3.8,0.23,0.25,0,30,degenaar2014,287.595,-5.7989,0.35,degenaar2014
XTE J1650-500,2001-2002,52149-52366,rxte,5.15,2.15,2.15,\citep{orosz2004},0.998,0,0,\citep{miller2002},Soft-to-hard transition with assumed mass,J,2.6,0.7,0.7,\citep{homan2006},4.38,0.26,0.34,47,89,orosz2004,252.5041,-49.9621,0.67,tomsick2004
XTE J1652-453,2009,55000-55158,rxte,none,none,none,none,0.13,0.29,0.29,\citep{chiang2012},Companion Star,E,5,0,100,\citep{reynolds2009},6.69,0.43,0.54,0,30,chiang2012,253.0847,-45.3444,6.7,hiemstra2011
XTE J1720-318,2003,52644-52832,rxte,none,none,none,none,none,none,none,none,Companion Star,E,6.5,3.5,3.5,\citep{chaty2006},2.98,1.19,1.65,0,89,none,259.9958,-31.7503,1.3, markwardt2003
XTE J1748-288,1998,50961 - 51038,rxte,none,none,none,none,none,none,none,none,HI Absorption,B,8,0,100,\citep{hjellming1998},8.66,1.98,1.84,,,,267.021083,-28.473833,Fitted (10.9),
XTE J1752-223,2009-2010,55091-55423,rxte,9.5,1.5,1.5,\citep{shaposhnikov2010},none,none,none,none,Hard-to-soft Transition (lower limit) \& Galactic Bulge (upper limit),H \& L,5.75,2.25,2.25,\citep{shaposhnikov2010,ratti2012},7.11,0.25,0.27,0,49,miller-jones2011,268.0629,-22.3423,0.5,ratti2012
XTE J1817-330,2006,53744-53996,rxte,none,none,none,none,none,none,none,none,Interstellar absorption \& Accretion rate,D \& L,5.5,4.5,4.5,\citep{sala2007},3.25,1.3,1.8,0,89,none,274.4314,-33.0188,0.12,rykoff2007
XTE J1818-245,2005,53575-53700,rxte,none,none,none,none,none,none,none,none,X-ray decay,F,3.55,0.75,0.75,\citep{cadolle2009},2.21,0.9,1.29,0,89,none,274.6018,-24.5383,0.54,cadolle2009
XTE J1859+226,1999-2000,51437-51661,rxte,>5.42,0,0,\citep{santana2011},none,none,none,none,X-ray Decay Time Scale \& Companion Star,F \& E,12.5,1.5,1.5,\citep{zurita2002,santana2011},4.13,1.65,2.29,0,70,corral-santana2011,284.6732,22.6582,0.6,{mrkwardt1999,dalfiume1999}
XTE J1908+094,2013-2014,56581-56780,maxi and xrt,none,none,none,,0.55,0.45,0.29,\citep{chiang2012},Companion Star,E,6.5,3.5,3.5,\citep{chaty2006},11.52,2.74,1.93,30,40,tao2015,287.2212,9.3847,Fitted,
\end{filecontents*}
\pgfplotstabletypeset[col sep=comma,
columns/Name/.style={string type,column name = Source,column type=l},columns/previous estimate method/.style={string type},columns/lit-dist-method-short/.style={string type,column name ={\makecell[t]{D$_{\text{lit}}$ Method$^a$}},string replace={none}{--}},columns/prevmethod-ref/.style={string type, column name={\makecell[t]{D$_{\text{lit}}$ Ref.}},string replace={none}{--}},columns/literaturemass-ref/.style={string type,column name = {\makecell[t]{M$_{\text{lit}}$ Ref.}},string replace={none}{--},string replace={}{--}},columns/literaturespin-ref/.style={string type,column name = a$_{\text{lit}}$ Ref.}, 
every head row/.style={before row=\toprule,after row=\midrule},
every last row/.style={after row=\bottomrule},
create on use/literaturemass/.style={
        create col/assign/.code={
            \edef\entry{\thisrow{literature-mass}$_{-\thisrow{literaturemass-lowerr}}^{+\thisrow{literaturmass-uperr}}$}
            \pgfkeyslet{/pgfplots/table/create col/next content}\entry}},
columns/literaturemass/.style={string type,column name={\makecell[t]{M$_{\text{lit}}$ \\ (\ms)}},string replace={$_{-}^{+}$}{ },string replace={none$_{-none}^{+none}$}{--},string replace={>5.42$_{-0}^{+0}$}{>5.42}
},
create on use/literaturespincombined/.style={
        create col/assign/.code={
            \edef\entry{\thisrow{literaturespin}$_{-\thisrow{literaturespin-lowerr}}^{+\thisrow{literaturespin-uperr}}$}
            \pgfkeyslet{/pgfplots/table/create col/next content}\entry}},
columns/literaturespincombined/.style={string type,column name=$a_{\text{lit}}$,string replace={$_{-}^{+}$}{ },string replace={none$_{-none}^{+none}$}{--},string replace={0.998$_{-0}^{+0}$}{$\sim0.998$}
},
create on use/distanceliterature/.style={
        create col/assign/.code={
            \edef\entry{\thisrow{prevmethod-dist}$_{-\thisrow{prevmethod-lowerr}}^{+\thisrow{prevmethod-uperr}}$}
            \pgfkeyslet{/pgfplots/table/create col/next content}\entry}},after row ={[0.6ex]},,
columns/distanceliterature/.style={string type,column name=D$_{\text{lit}}$ (kpc),string replace={$_{-}^{+}$}{},string replace={none$_{-none}^{+none}$}{--},string replace={7$_{-0}^{+0}$}{$\sim7$},string replace={<8.5$_{-}^{+}$}{$<8.5$},string replace={5.9$_{-0}^{+100}$}{$>5.9$},string replace={8.5$_{-8.5}^{+0}$}{$>8.5$},string replace={6$_{-0}^{+100}$}{$>6$},string replace={1.5$_{-0}^{+0}$}{$\sim1.5$},string replace={5$_{-0}^{+100}$}{$>5$},string replace={8$_{-0}^{+100}$}{$>8$},string replace={8.5$_{-0}^{+100}$}{$>8.5$}},
create on use/bestdistance/.style={
        create col/assign/.code={
            \edef\entry{\thisrow{best-estimate-med}$_{-\thisrow{best-estimate-low}}^{+\thisrow{best-estimate-upper}}$}
            \pgfkeyslet{/pgfplots/table/create col/next content}\entry}},after row ={[0.6ex]},
,columns/bestdistance/.style={string type,column name={\makecell[t]{Best Estimated D from this work (kpc)$^b$}},string replace={$_{-}^{+}$}{},string replace={none$_{-none}^{+none}$}{--},string replace={$_{-0}^{+0}$}{},string replace={<8.5$_{-}^{+}$}{$<8.5$},string replace={>5.9$_{-}^{+}$}{$>5.9$}}
,columns={Name,bestdistance,distanceliterature,prevmethod-ref,lit-dist-method-short}
]{tables/compTable.csv}
\begin{tablenotes}[flushleft]
\small
\item \textbf{Notes.}
\item $^{a}$ A: Gaia parallax, B: H\,{\sc i} absorption, C: Reflection \& continuum-fitting, D: Interstellar extinction, E: Companion star, F: X-ray decay time scale, G: X-ray luminosity \& disk temperature, H: Hard-to-soft transition luminosity, I: Proper Motion (Jet), J: Soft state and soft-to-hard transition without corrections, K: Radio \& X-ray correlation, L: Assumed to be at Galactic Bulge. M: Flux Scaling, O: Soft state \& peak luminosity.\\
\item $^{b}$ Quoted errors are statistical (1$\sigma$). Refer to Section~\ref{sec:limitations} for discussion of systematic errors.
\end{tablenotes}
\end{table*}

\bsp	
\label{lastpage}
\end{document}